# Unbiased Library of *k*-regular, *n*-sized, Connected, Small Graphs

Tamas David-Barrett

University of Oxford

Email: tamas.david-barrett@trinity.ox.ac.uk

Address: Trinity College, Broad Street, Oxford, OX1 3BH, UK

Web: www.tamasdavidbarrett.com

## Abstract

The past decade highlighted the usefulness of social network simulations that run on k-regular, n-size, connected graphs. These can be seen as small-scale models of human social networks of large societies. By narrowing down onto k-regular graphs, the degree variation can be eliminated from the research question, which allows a focus on the isolated impact by other variables, for instance, by the clustering coefficient or the size of the network. This paper describes the generation of a random graph library that uses a random walk graph creation algorithm that starts from the 'chain of caves', which is the structure in which the clustering coefficient is at its maximum. This method finds mid and high clustering coefficient graphs, while Wolfram's RandomGraph was useful for finding low ones. The merge of the two samples proved to be somewhat biased. After eliminating a host of network measures, the paper focused on mean graph distance as a further variable and created an unbiased subsample for each size and clustering coefficient bin.

# Introduction

The past decade has seen the rise of studying network behaviours emerging on a subset of graphs: $k$-regular, $n$-sized, connected graphs, that is, networks in which there are $n$ agents, each agents has the same $k$ edges, and there are no separated subgraphs (1, 2). These graphs offer an easy way of modelling network processes of large-scale human societies on small, computationally accessible way.

The 'trick' of downsizing the modelling scale is possible due to two assumptions: (i) the focal social networks are regular, i.e., every node is connected the same number of others, and (ii) that $k$ is much smaller than $n$, i.e., these graphs are sparse. These assumptions narrow down the possible graph structures and eliminate the degree $k$ as a variable from the models, making simulations both computationally scalable and the models trackable.

These two assumptions about human social networks are not entirely unrealistic. Although there is certainly variation in the number of social network partners that humans have, the number of meaningful relationships is in a relatively narrow range compared to the size of the large societies most human live in today. For instance, a superstar musician may have billions of fans, they might have even personally played music directly to millions of concert goers, yet, their number of actual social connections with meaningful emotional depth is likely to be nearer to a couple of hundred, and certainly under a thousand (3). At the other end of the scale, almost no human has zero social relations. The number of social relationships people have is in a relatively small range, and thus, from the point of view of a large society that consists many millions, it is a reasonable modelling assumption that everyone has the same, small number of connections. This assumption allows the elimination of $k$ as a moving part of these models, which in turn allows isolating the effect of other variables.

There are many social phenomena in today's human societies, which appear on the scale of the society but are generated by behaviour on the individuals' level. For instance, the effect of demographic processes on social trust, changing cooperative behaviour, and the shift away from kinship to friendship as the main social network building behaviour (4, 5). These are inherently difficult to model, because closed form solutions do not exist, and simulating excessively large networks is forbiddingly expensive even at today's computer speed.

It is at this computational constraint where the assumption that everyone has the same number of connections helps. This assumption opens the possibility to move the focus to other variables, e.g., the effect of group size or density, while the number of connections stays fixed. This assumption has been useful in understanding a wide range of phenomena, from the social brain hypothesis (6), to speed of the evolution of linguistic communication (7), to the evolutionary foundations of religiosity (8) and of social stratification (9), to the origin of mass ideologies (10), to rise of modern friendship (5).

Many of these papers used their own graph libraries for the specific purpose of their focus. These were all either generated by a model via an edge switching algorithm particular to the paper's dynamics or were generated randomly by the mathematical software used.

For future research, it is useful to create a standardised library of graphs that allows both that several papers use the same unbiased sample of graph structures, and that the *exact* replication of the calculations will be possible. Due to recent results that pinpoint the groups size and the clustering coefficient as key variables behind the cooperativeness of a social network graph (11), the focus of this quest is an unbiased sample that covers the entire theoretically possible clustering coefficient range for small to medium size graphs. This paper's goal is to create and document such a graph library.

# Methods

The objective is to generate graph libraries for different group sizes, such that the elements of the libraries represent the entire range of possible clustering coefficient values. Note that the possible number of isomorphic networks is extremely high for most relevant group sizes, and some are not currently known (12, 13). Because of this fact, I had to create the graph libraries taking samples.

For this, I selected the degree, $k$, to be fixed at 4 for all nodes. Choosing the lowest possible $k$ is beneficial for then the $n$ can stay relatively small, which is computationally cheaper. At the same time, many limiting graph behaviours, like excessively slow or no synchronisation, come in at k<3, and k=3 limits the possible number of clustering coefficient positions inside the range to the point that is meaningless.

The number of 4-regular 10-sized non-isomorphic graphs is known, 59, which are all in graph library of $n$=10. In this case, there was no need to sample.

To generate the graph libraries for the network sizes for which the number of possible graphs is not known, first I had to have an estimate for the total size of the possible sets of graphs.

The number of graph structures of k-regular graphs increases with $n$ faster than exponential (Fig. 1). To my knowledge, the theoretical size of the graph library of $k$-regular graphs is not known for $n$=20, 25, 30, …. However, a log-linear transformation of the known graph sizes from $n$=6 to 18 yields the predicted number of non-isomorphic graphs = $10^{(6.77 + 1.56 \ x - 8.93 \ \text{Log}_{10} \ x)}$ at AR2=99.95% AR2. This suggests, for instance, that the set size for $n$=20 and 30 are in the magnitudes of $10^{11}$ and $10^{23}$ respectively (Fig. 1).

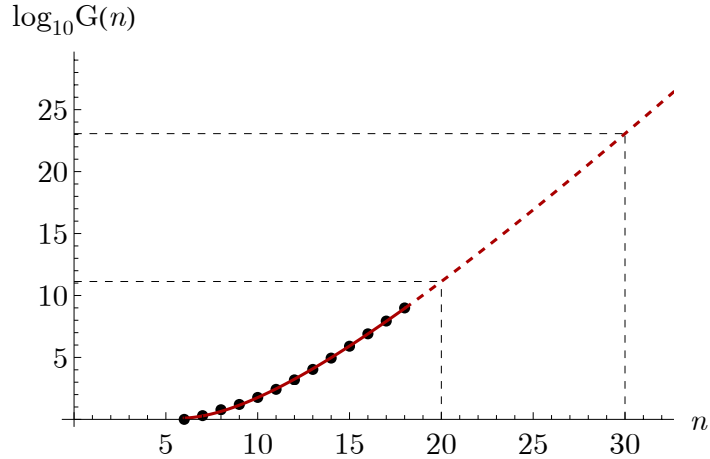

Fig. 1. The size of the set that contains all possible $n$-sized 4-regular graphs, G($n$) grows increases faster than exponential.

Given these extremely large G($n$) sets, it was both impractical and impossible to sample from a fully known set for the range of graph sizes for which the scientific questions are relevant. Because we are interested in how the clustering coefficient affects the graph-level phenomena, e.g., coordination efficiency, this sample would need to be informed by the range of clustering coefficient among all possible graphs, and then select a consistent way to find them.

First, the clustering coefficient range. Although, it is not practical to know all elements of the graph sets of size in the millions and above, it is possible to calculate the range of the clustering coefficient (14). In general, it is always possible to construct a graph with zero triangles as long as $k$ is much lower than *n*, while the upper limit of the clustering coefficient depends only on the number of edges each node has, *k*, and not the size of the graph, *n*:

$$\chi \leq 1 - \frac{6}{k(k+1)}$$

for any *k*-regular connected graph (14). For *k*=4, which is the model parameter for the Microfoundations theory papers (see discussion above), this yields $\chi = 0.7$, which is indeed the maximum clustering coefficient of the graphs in the graph library for *n*=10 (Fig. 2).

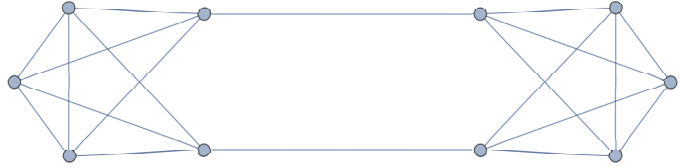

Fig. 2. The graph with maximum clustering coefficient ($\chi = 0.7$) for *n*=10 and *k*=4.

This structure can be infinitely extended as long as one of the subgraph 'halves' is added in a chain, in which case the above inequality is always equality (14). In fact, this is a version of the original 'caveman' graph structure (15), modified for regular graphs. Each 'cave' holds exactly *k*+1 cave nodes, and hence that any number of cave-chain extensions are possible if *n* is a multiple of *k*+1. This way it works also for *n*=20, 25, 30, and so on for *k*=4. In short, the clustering coefficient's range is 0 to 0.7 independent of *n*.

The second problem is finding the graphs that correspond to the different clustering coefficient levels. One possibility is to use standard random graph generators, for instance, Mathematica's RandomGraph function used together with the DegreeGraphDistrubution option (abbreviated as WM). This generator is excellent to find low and lower mid-range clustering coefficient graphs, but not high ones when $n$ is not small, and hence leave about half the range empty for medium sized graphs.

To generate the high clustering coefficient graph samples, I used an edge swap algorithm (described below, and abbreviated as CC). Each run started with the relevant version of the 0.7 clustering coefficient graph, i.e., the one with the cave chain (Fig. 3). Then I applied an edge swap algorithm sequentially, recording the graphs at each step. This algorithm is similar in spirit to the random-walk-based graph generators in the literature (16).

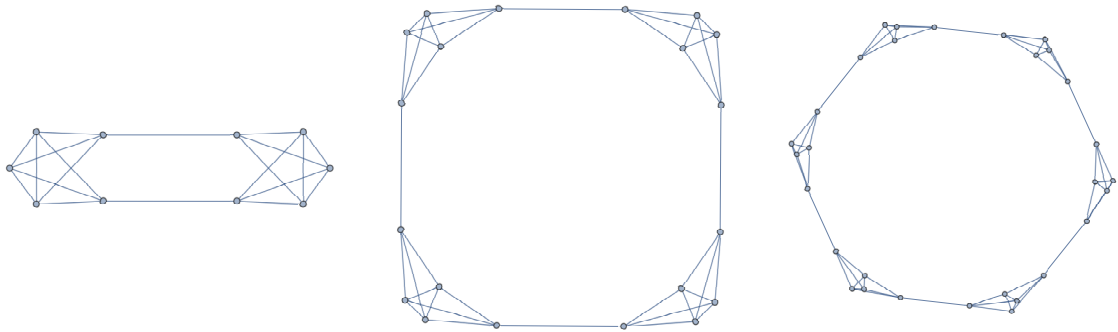

Fig. 3. The 'cave-chain' starting points for the 'build-down' random graph generating algorithm. Sizes, from left to right: n=10, 20, 30.

I used the following 'build-down' algorithm.

1. Select node $x_1$ randomly.

2. Select node $x_2$ randomly among the alters of $x_1$.

3. Select node $x_3$ randomly among the vertices that are neither connected to $x_1$ nor $x_3$.

4. Select node $x_4$ randomly among the vertices that are not $x_2$ while also not connected to $x_3$.

After the vertices are selected, I used the following algorithms for edge swapping:

1. Delete the edges $x_1$- $x_2$ and $x_3$ - $x_4$ (if they exist)

2. Create the edges $x_1$- $x_3$ and $x_2$ - $x_4$ (if they did not exist)

3. Check if the degree is still universally $k$. Abort if not.

4. Check if the graph is still connected. Abort if not.

5. Check that none-of the new edges are self-loops. Abort if either is.

6. Check if the generated graph is isomorphic with any of the other graphs in this particular batch of graphs. Abort if yes.

7. Measure the average clustering coefficient, add graph to the relevant clustering coefficient bin.

8. Run until either the number of graphs in the clustering coefficient bin reaches 20, or the number of consequent aborted runs reaches 500. (The presence of a limit per bin per run reduces the distortion among the algorithms.)

9. Add the collected batch of graphs to the raw sample.

Using a combination of Mathematica's random graph generator (WM), and the sequential 'build-down' algorithms from the cave-chain graph (CC), I created a 'raw sample' of graphs for each of $n$=15 to 50 in 5-step increments for $k$=4, made up of bins that correspond to the theoretically possible clustering coefficients. (NB. As far as I know, the CC algorithm is new in the literature for generating regular graphs randomly.)

The results show that the two different algorithms are good at finding different parts of the G($n$) graph populations. The WM algo tends to find low $\chi$ graphs but not high ones, and the CC algo finds mid and high $\chi$ graphs but not low ones (see Fig. 4).

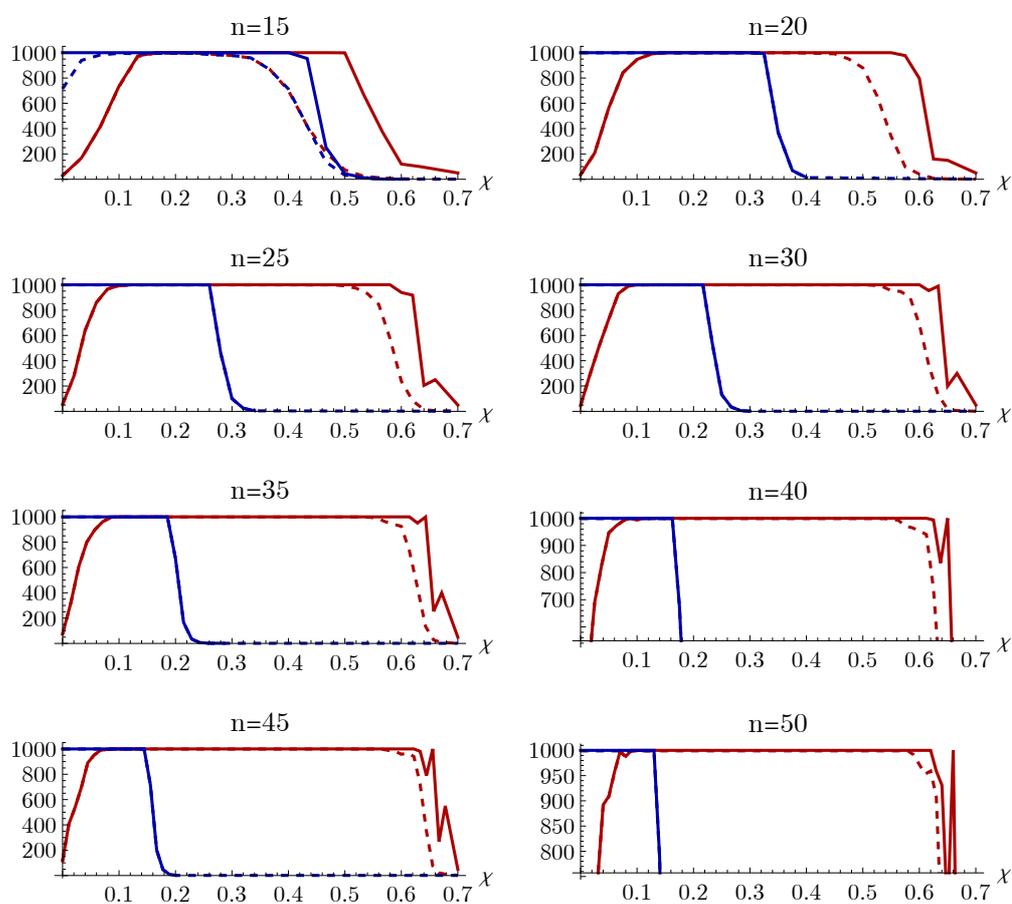

Fig. 4. Histogram of sample sizes. Blue: MW algorithm, red: CC algorithm. Continuous line: raw sample, dashed line: non-isomorphic sample.

Notice that the difference in between the two algorithms, depicted in Fig. 4, may bring a bias into calculating the effect of the clustering coefficient on graph-level behaviour. For instance, if the behaviour is dependent on a structural property other than the level of clustering coefficient for graphs more likely to be found by the WM algo than the CC algo then $\chi$-based calculation might pick up the effect of the sampling algorithm rather than the $\chi$. To check for this possibility, we need to select the alternative variables.

It is not immediately clear which variables to focus on among the wide range of possible graph measurements. In other words, which variable to select as most likely to distort the clustering coefficient focused results. This question would be only decidable based on the scientific question we would ask in our calculations. Without a modelling question, it is not immediately obvious which variables to focus on. Let us here consider the following variables (apart from the clustering coefficient), which are all frequently used in network science: eigenvector centrality, page rank, graph distance, vertex betweenness, edge betweenness, and closeness centrality.

It turns out that it is possible to narrow down this variable list:

1. There are graph measures, e.g., eigenvector centrality (and thus page rank centrality), that are useful network science tools comparing vertices, but as graph average they are determined by simply the size of the graph, and hence do not vary at all. NB. This holds for any graph, not only fixed degree ones, or only for ones for which the total number of edges is constant.

2. Some measures have perfect correlation with each other. For instance, the average graph distance among the vertices is perfectly correlated with the average vertex betweenness centrality. In this case, it is enough to focus on a single variable, for instance, on the graph distance. (Choosing the graph distance to focus on is somewhat arbitrary, although may be justified by it being the most immediately meaningful for the scientific question that motivated this paper, as it provides insight into how it affects the speed of synchronisation.) NB. Just as above, this is also true for all graphs, with fixed $n$.

3. Some measures correlate with the average graph distance when the total number of edges is fixed. For instance, average graph distance correlates perfectly with the average edge betweenness centrality as long as both $n$ and the total number of edges are fixed. NB. This means that the correspondence is the same for fixed degree graphs with $k=4$ and random graphs for which the number of edges is $\frac{1}{2}kn$.

4. The average closeness centrality shows relatively loose correlation with the average graph distance for general random graphs. However, for fixed degree graphs this becomes perfect or near perfect correlation (0.98 to 1) depending on the clustering coefficient.

As a consequence, it is justified to focus on the average graph distance as the main variable in which we need to test the different B($n$, $\chi$) bins.

To see whether the WM and CC algorithms differ in the kinds of graphs they find we need comparable subsamples. Unfortunately, the size of the samples found by the WM and CC methods varies across the $\chi$ range (Fig. 4). Fortunately, however, there is substantial overlap between WM and CC in terms of what range of the clustering coefficient they both found, and thus it is possible to test if the graph distance means of the two sub-samples are the same. For this, for each $n$, I selected from the $\chi$-bins for which both algorithms found at least 990 non-isomorphic graphs. I combined these bins to depict the histograms of the mean graph distance. The results shows that the two algorithms find somewhat different subsamples (Fig. 5).

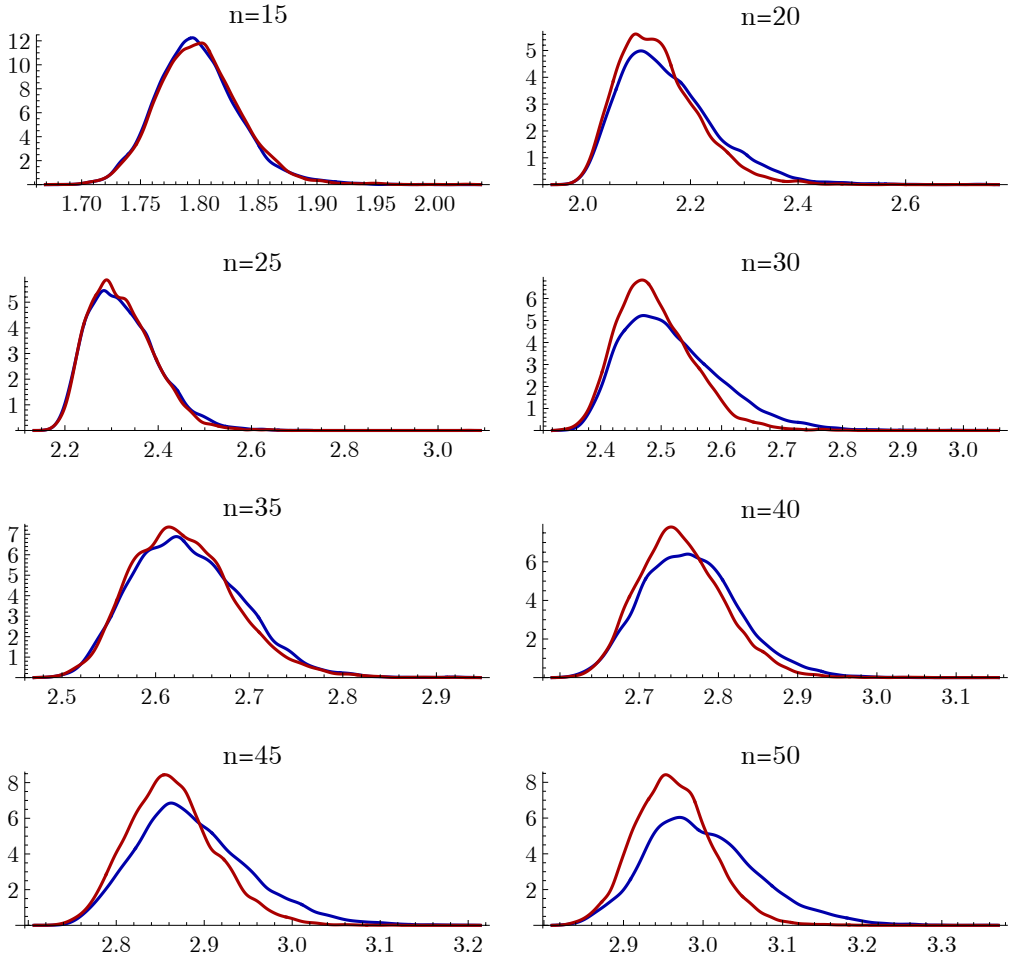

Fig. 5. Histogram of mean graph distances in the $\chi$ bins for which both methods found at least 990 out of 1000, non-isomorphic graphs. Blue: WM algorithm, red: CC algorithm.

Notice that the difference in the distributions in the Fig. 5 histograms is not substantial, it might be enough to confound modelling results. Such distortion would happen, for instance, if a graph's behaviour was driven by the mean graph distance and not by the clustering coefficient. As the CC algorithm finds graphs with somewhat lower mean graph distance than the WM algorithm for large graphs, while CC is better at finding high clustering coefficient graphs than WM, a graph-level behaviour that might seem the consequence of the clustering coefficient could be driven instead by the graph distance difference between the two algorithms.

The group size and the clustering coefficient might not be the only graph traits that affect a behaviour. For instance, while the clustering coefficient is likely to be the only trait that is important for the rise of costly cooperation on reputation networks (17), a behavioural synchrony algorithm (e.g., 2) will be dependent on the average path length even when both $n$ and $\chi$ are fixed. A model of the trade-off between these two behaviours (i.e., strategic cooperation vs behavioural synchrony) could be biased by the fact that the low $\chi$ sample bins would be found by the WM algorithm that tends to find graphs with somewhat higher graph distance, and vice versa for the CC algorithm.

Unfortunately, there is not much we can do about this possible bias, apart from being aware of the distortion, as well as robustness testing the findings of any particular graph-behaviour model for both algorithms. (NB. The distribution for the full G($n$) populations in terms of different graph traits, is not known.) Notice, however, that the difference is not very large between the respective distributions. Hence, one possible modelling rule could be that the clustering coefficient needs to have a large effect size as a variable, which would make it less likely that the confounding mechanism was at play.

Given this proviso, notice that the two distributions raise a further issue: they are skewed. This raises a potential further bias. If the skew varies across the $\chi$ range, then this in itself can create a similar confounding effect as described above. The solution may be to select a normally distributed sub-sample. For this, I created a library of final $B_F(n, \chi)$ bins with up to 100 graphs in each. For sampling, I used the following algorithm:

Step 1.    For each $n$ and $\chi$, I merged the CC and WM non-isomorphic samples. This created merged bins for each $\{n, \chi\}$ pair

Step 2.    I ran a structure detection algorithm to eliminate the isomorphic graphs from the merged bins. This created merged non-isomorphic bins.

Only the $n$=15 sample contained overlap between the two algorithms (Fig. 6). For all other $n$s, the samples were entire non-isomorphic.

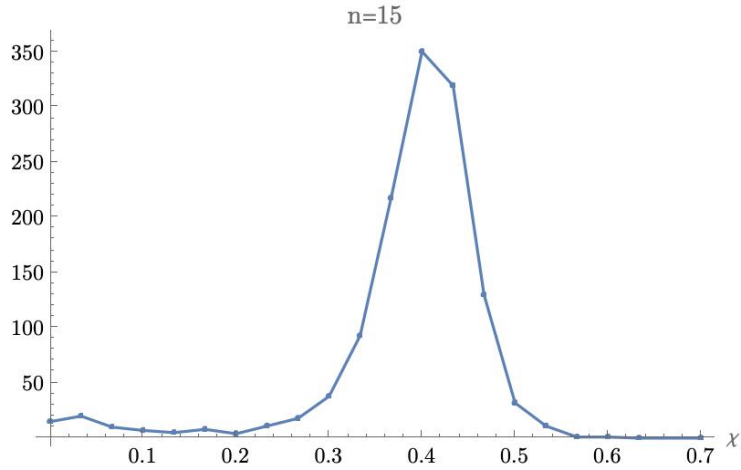

Fig. 6. Number of isomorphic graphs between the WM and CC algorithms.

Step 3.    I ran a search algorithm on the merged non-isometric bins aiming to find a batch of 100 graphs that we are normally distributed in graph distance. For this, randomly drew batches of 100 graphs from the merged non-isometric sample. For each batch, I calculated the Cramer-von-Mises test of normality. I kept the sample with the highest test score. I ran this for 10k times. If at that point the highest test score was above 0.999, then I kept this batch. If it did not, then I ran the draws until either the limit was reached, and the number of draws reach 100k. This yielded the final sample. (NB. When the merged non-isomorphic sample bin size is 100 or lower, the final bin is the same.)

The normality test histograms across all bins show that the above method yielded an almost entirely normally distributed sample (Fig. 7). (Fig. S1 depicts the graph distance distributions in both the merged non-isomorphic sample and the final sample for each bin.)

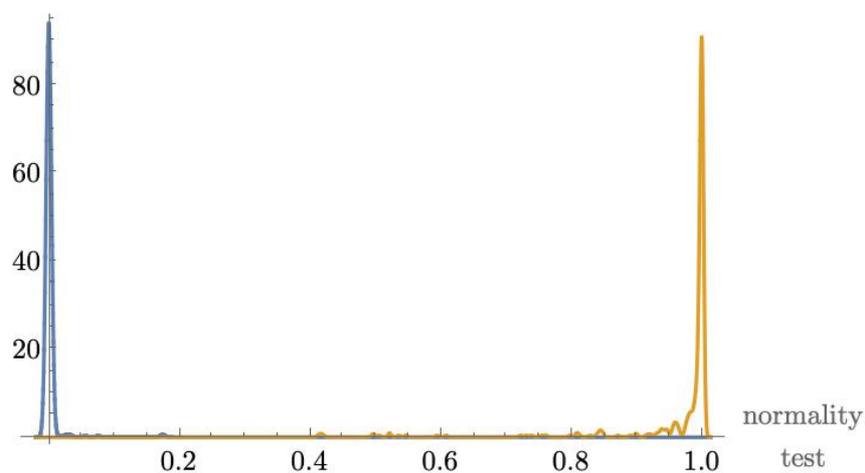

Fig. 7. Histogram of Cramer-von-Mises normality tests across all ns χs. Blue: merged non-isomorphic sample, orange: final sample.

It is important that the above method of creating the final sample does not introduce a new bias. Fig. 8 shows the relationship between the clustering coefficient and the means of the average graph distance for each bin. There is no systemic deviation in the final sample compared to the merged non-isometric sample.

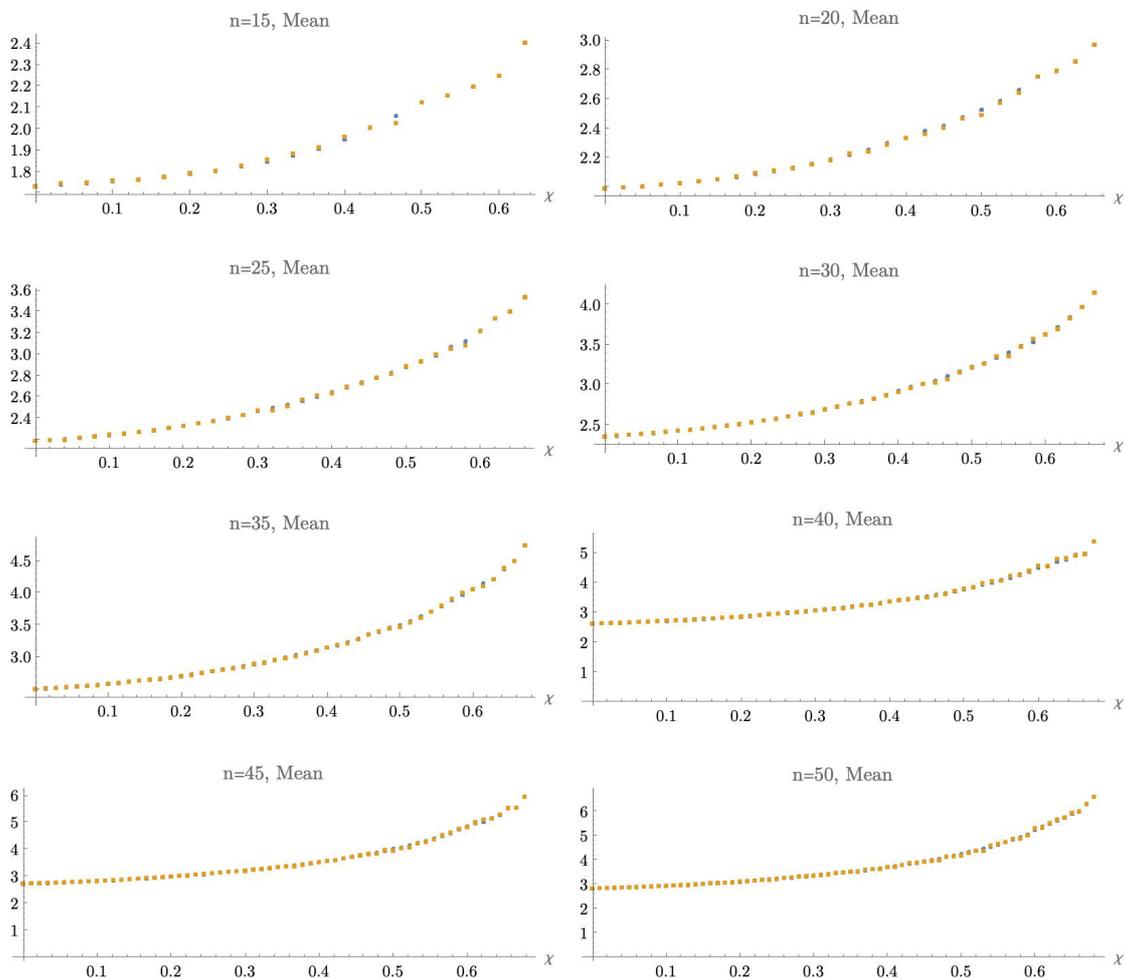

Fig. 8. The mean of average graph distances in each bin. Blue: merged non-isomorphic sample, orange: final sample.

This way of choosing a normally distributed sample came at the cost of somewhat reduced variation in each bin (Fig. 9).

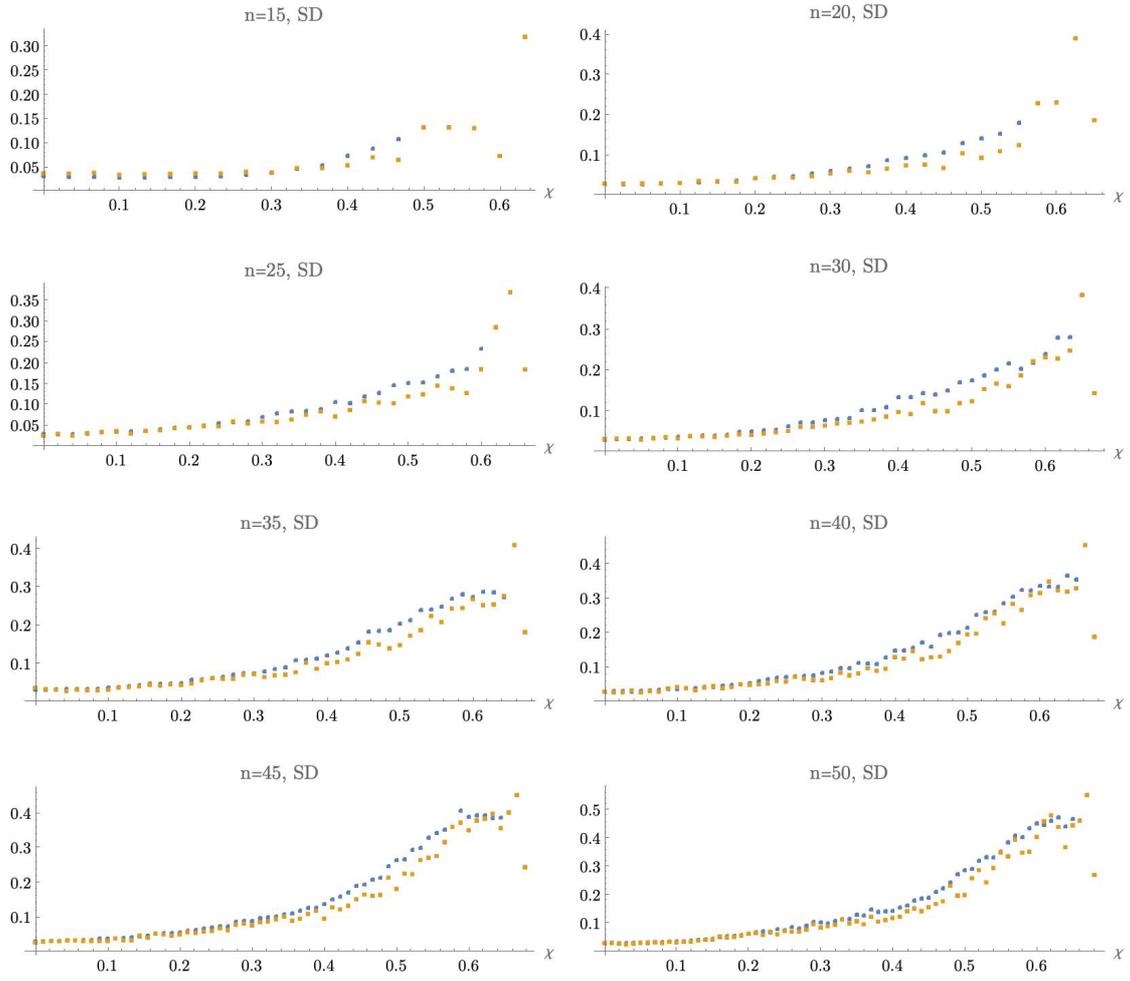

Fig. 9. The standard deviation of average graph distances in each bin. Blue: merged non-isomorphic sample, orange: final sample.

The skewness is substantially reduced (Fig. 10).

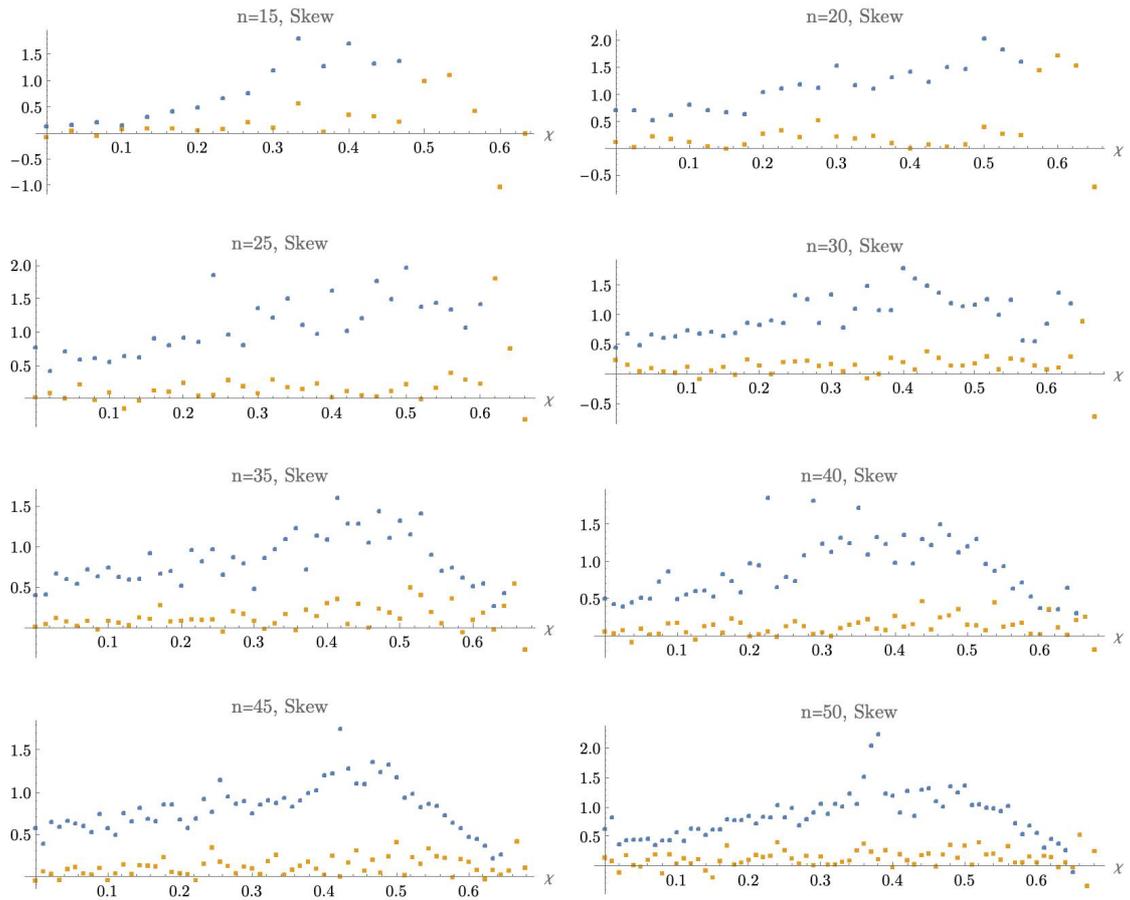

Fig. 10. The skewness of average graph distances in each bin. Blue: merged non-isomorphic sample, orange: final sample.

The final sample is normally distributed for all n>15, and near normally distributed for n=15 (Fig. 11).

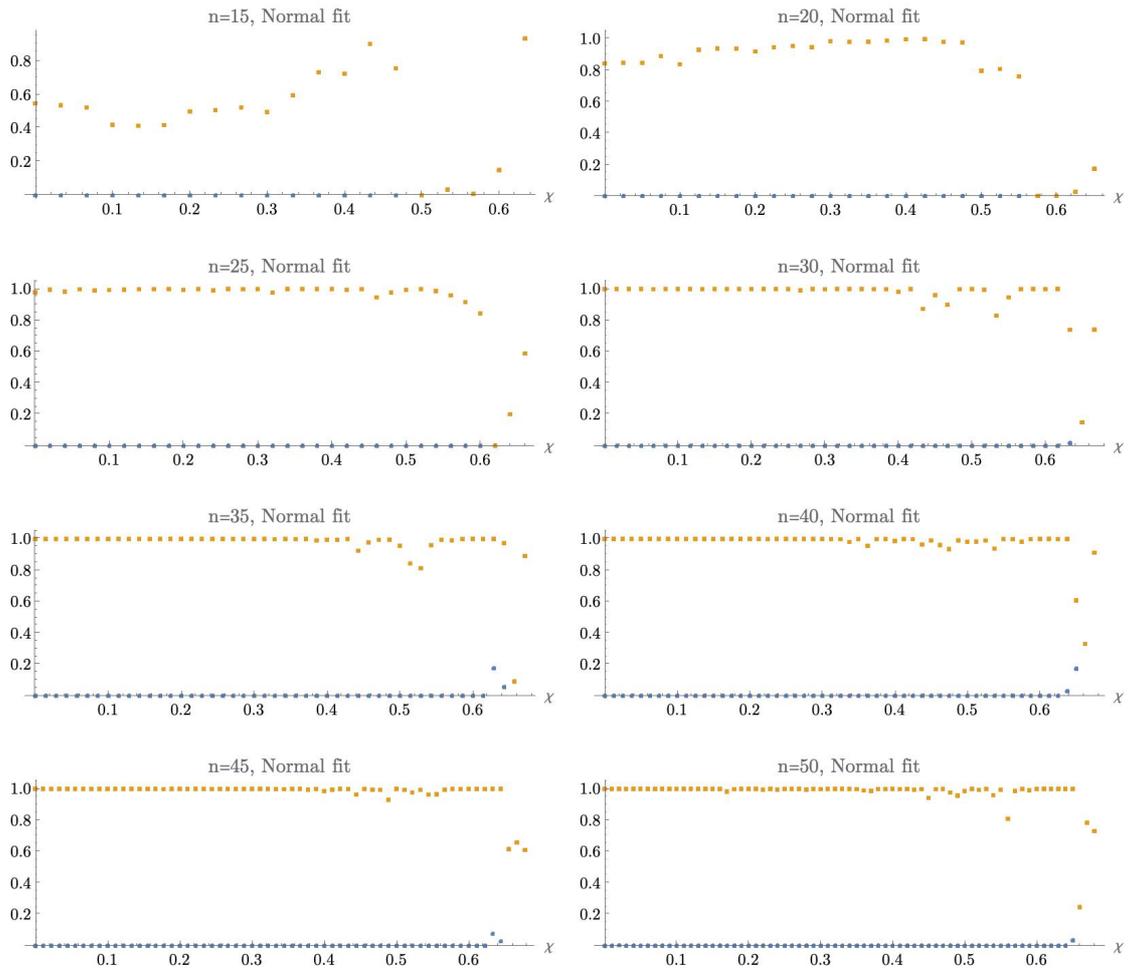

Fig. 11. Cramer-von-Mises normality test of average graph distances in each bin. Blue: merged non-isomorphic sample, orange: final sample.

For the overview sample size statistics, see Table 1.

**Table 1.** Sample size statistics. The term 'built-in' refers to the number graphs in the raw sample using the random graph generator of Mathematica. Similarly, the terms 'algo 1', 'algo2', and 'algo3' refer the different versions of the 'build-down' algorithms defined the paper. Library size refers to the number of graphs in the final sample, in which each bin has 100 elements, except for the cases in which either only fewer exist, or the algorithms found only fewer than 100 cases.

| n | $\chi$ bins | Number of all graphs | Method | WM raw | WM non-iso | CC raw | CC non-iso | Merged non-iso | Final library size |
|---|---|---|---|---|---|---|---|---|---|
| 10 | 12 | 59 | All found | | | | | | 59 |
| 15 | 21 | 805,491 | Sampled | 14,264 | 12,708 | 14,634 | 10,540 | 21,954 | 1,615 |
| 20 | 28 | $> 10^{11}$ | Sampled | 14,445 | 14,429 | 22,722 | 19,534 | 33,692 | 2,450 |
| 25 | 35 | $> 10^{16}$ | Sampled | 14,592 | 14,575 | 30,156 | 27,423 | 41,997 | 3,204 |
| 30 | 42 | $> 10^{23}$ | Sampled | 14,713 | 14,695 | 36,997 | 34,470 | 49,164 | 3,925 |
| 35 | 49 | $> 10^{29}$ | Sampled | 14,884 | 14,866 | 44,307 | 41,774 | 56,639 | 4,636 |
| 40 | 56 | $> 10^{36}$ | Sampled | 14,890 | 14,873 | 51,492 | 48,985 | 63,857 | 5,326 |
| 45 | 63 | $> 10^{42}$ | Sampled | 14,979 | 14,961 | 58,225 | 55,545 | 70,505 | 6,009 |
| 50 | 70 | $> 10^{49}$ | Sampled | 15,100 | 15,082 | 65,032 | 62,245 | 77,326 | 6,700 |

# Results and Discussion

The resulting graph library is available at:

www.tamasdavidbarrett.com/s/data_for_unbiased_k-regular_n-sized_graphs.zip.

These graphs may be useful for studying group coordination, and the impact of low-scale, individual-level behaviour on the spread of behavioural norms and culture in large societies.

# Supplementary Material

Fig. S1 below contains the distribution histograms of the average graph distance for each $\{n, \chi\}$ bin. Blue: merged non-isomorphic sample, orange: final sample. The bin size refers to the number of graphs in the relevant bin of the merged non-isomorphic sample. The tests are the Cramer-von-Mises normality test for the two samples. (Only the bins larger than 100 graphs were depicted, as when the merged non-isomorphic sample bin size is 100 or lower, the final bin is the same.)

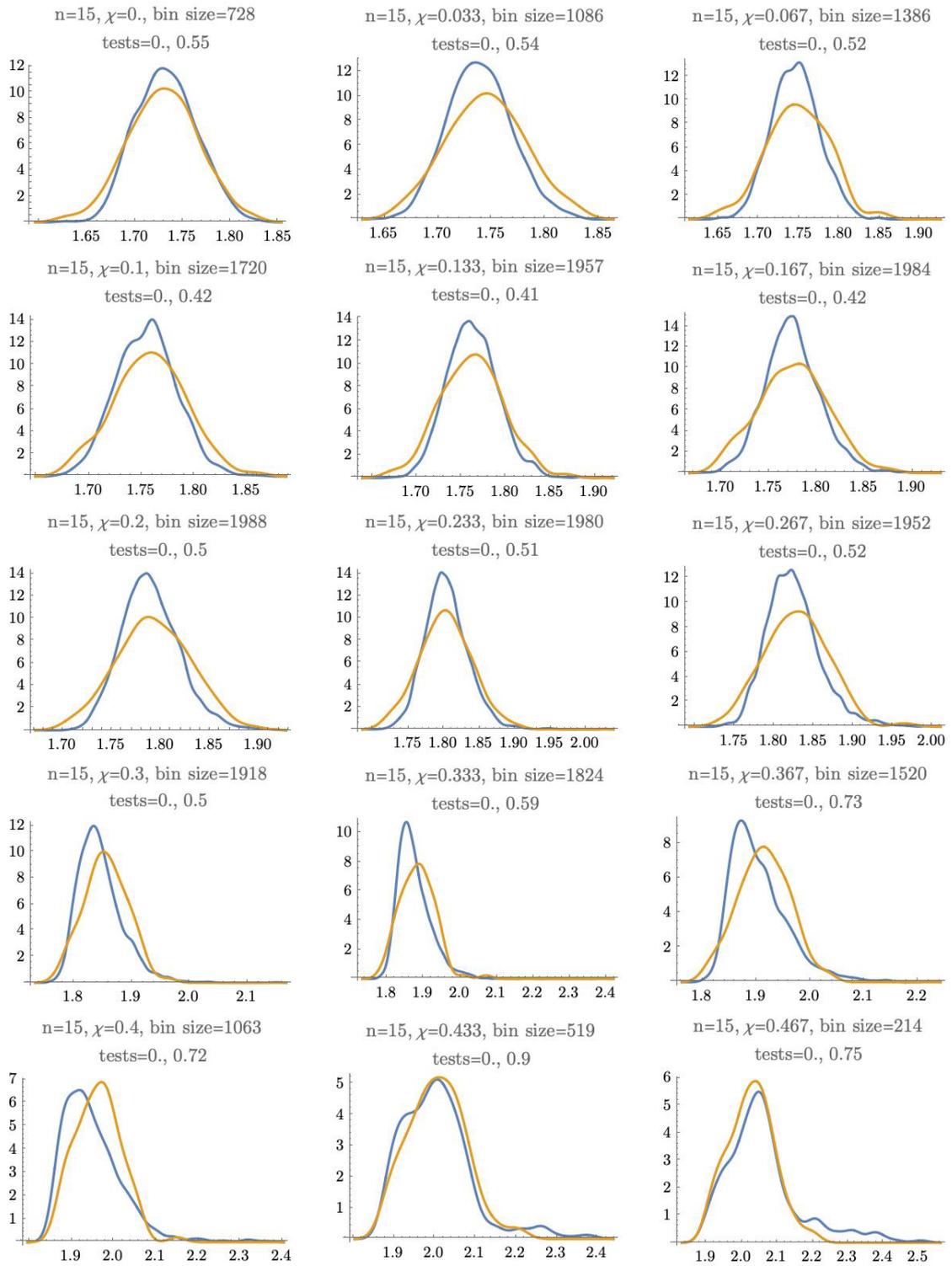

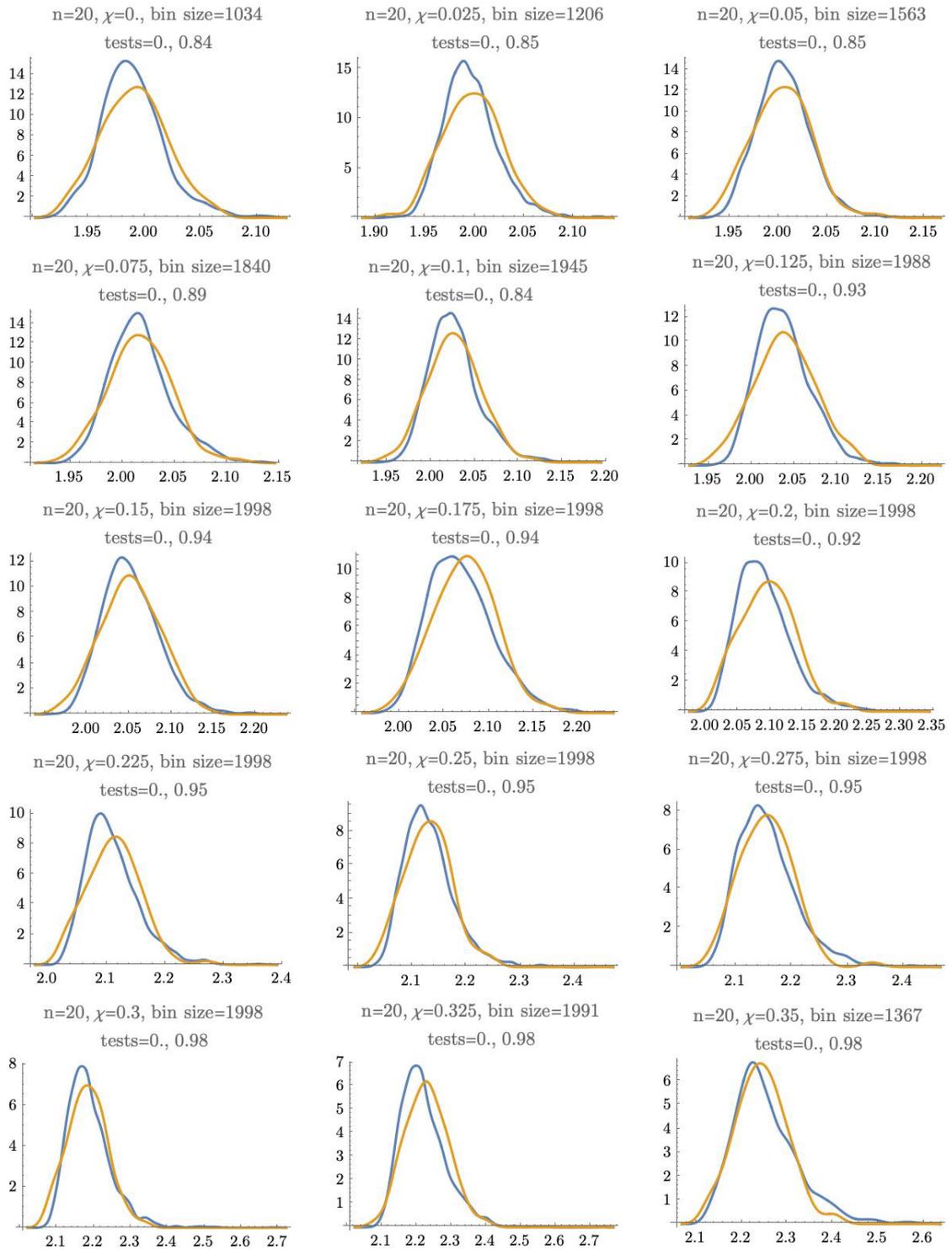

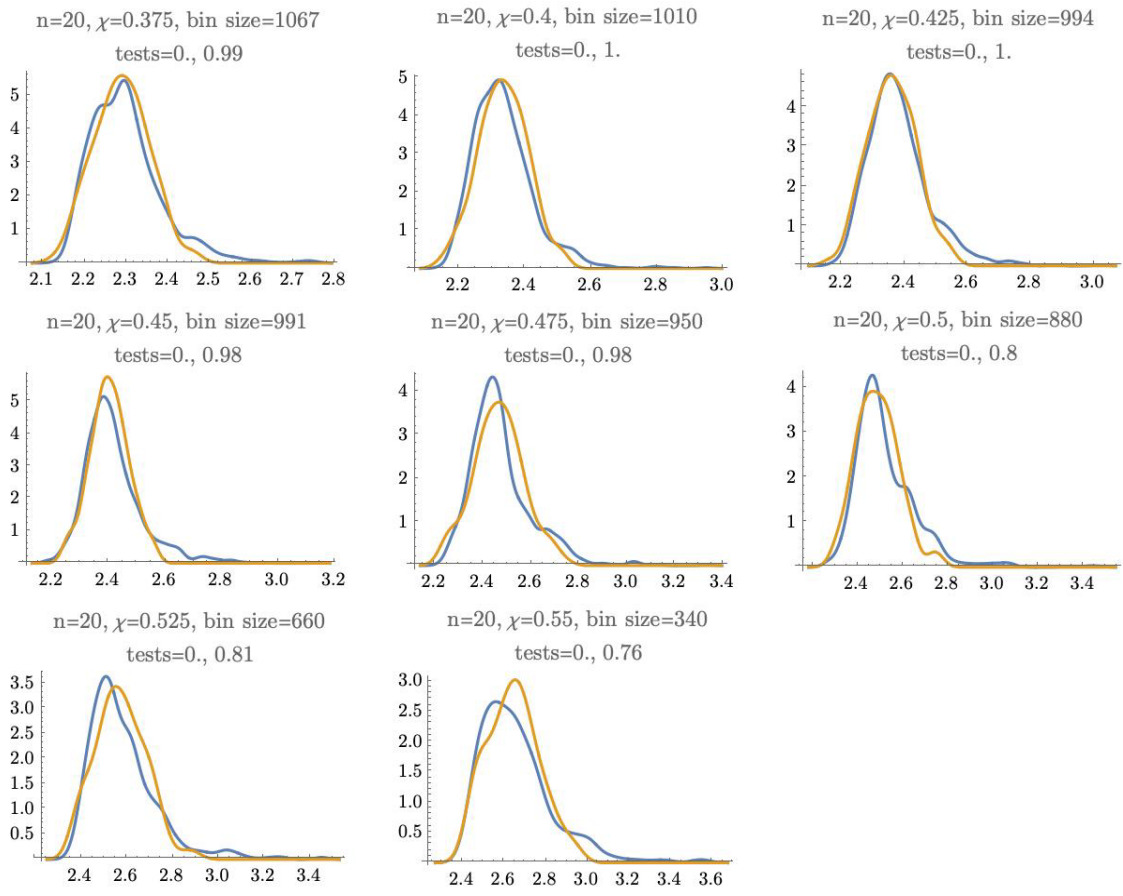

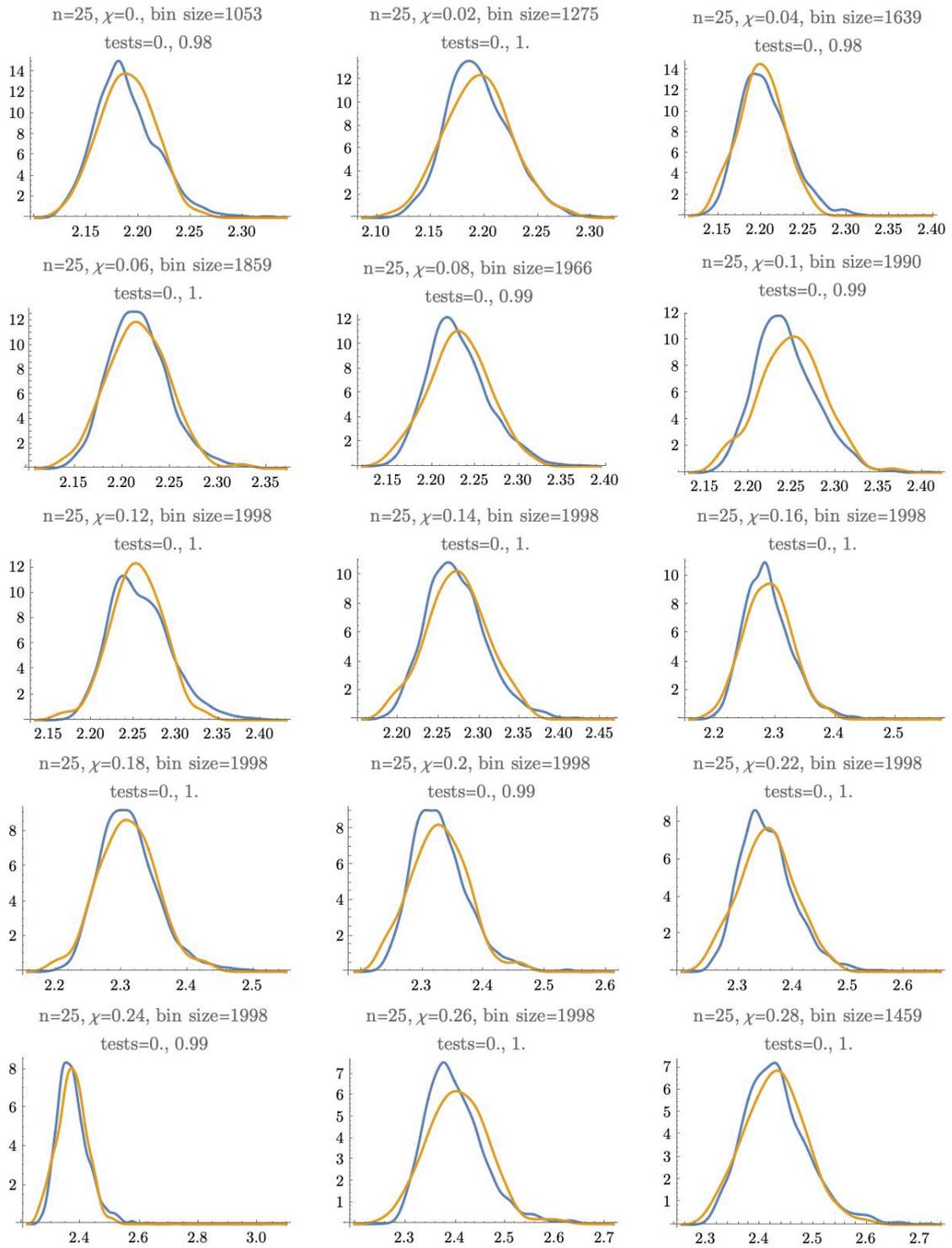

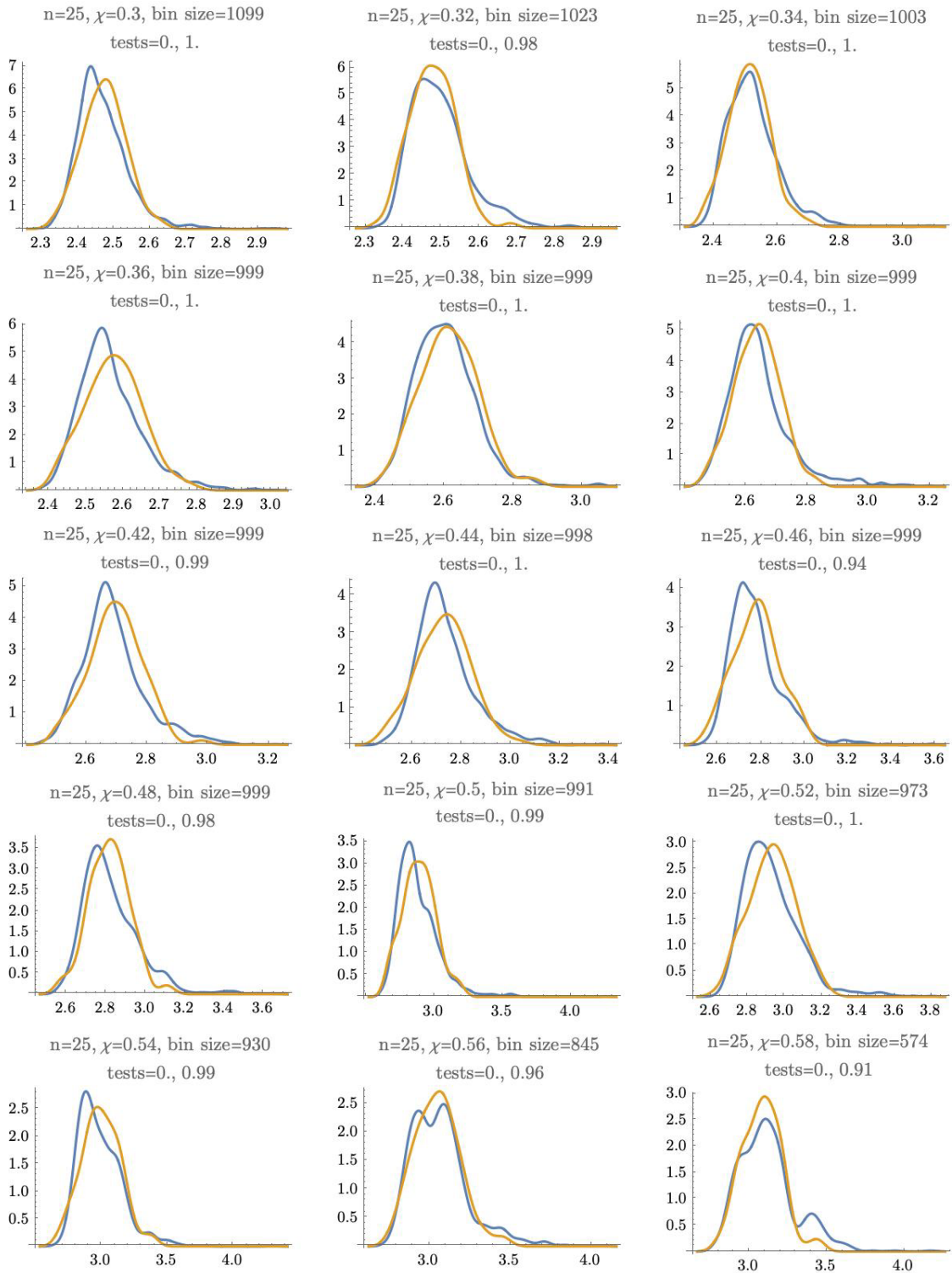

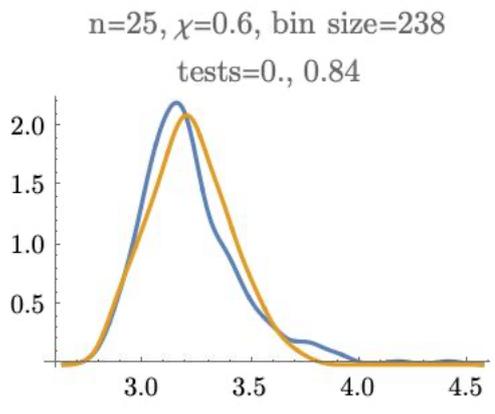

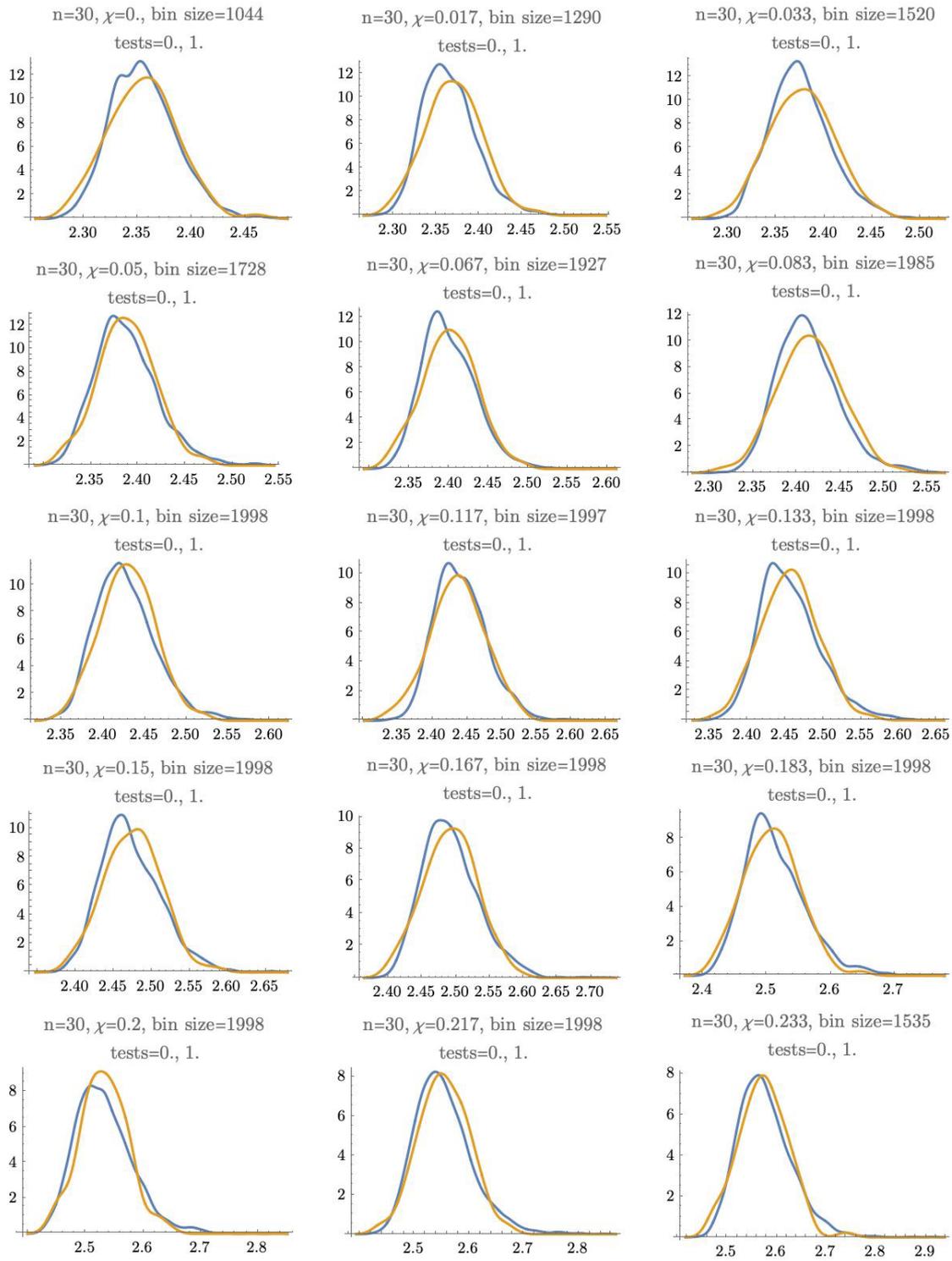

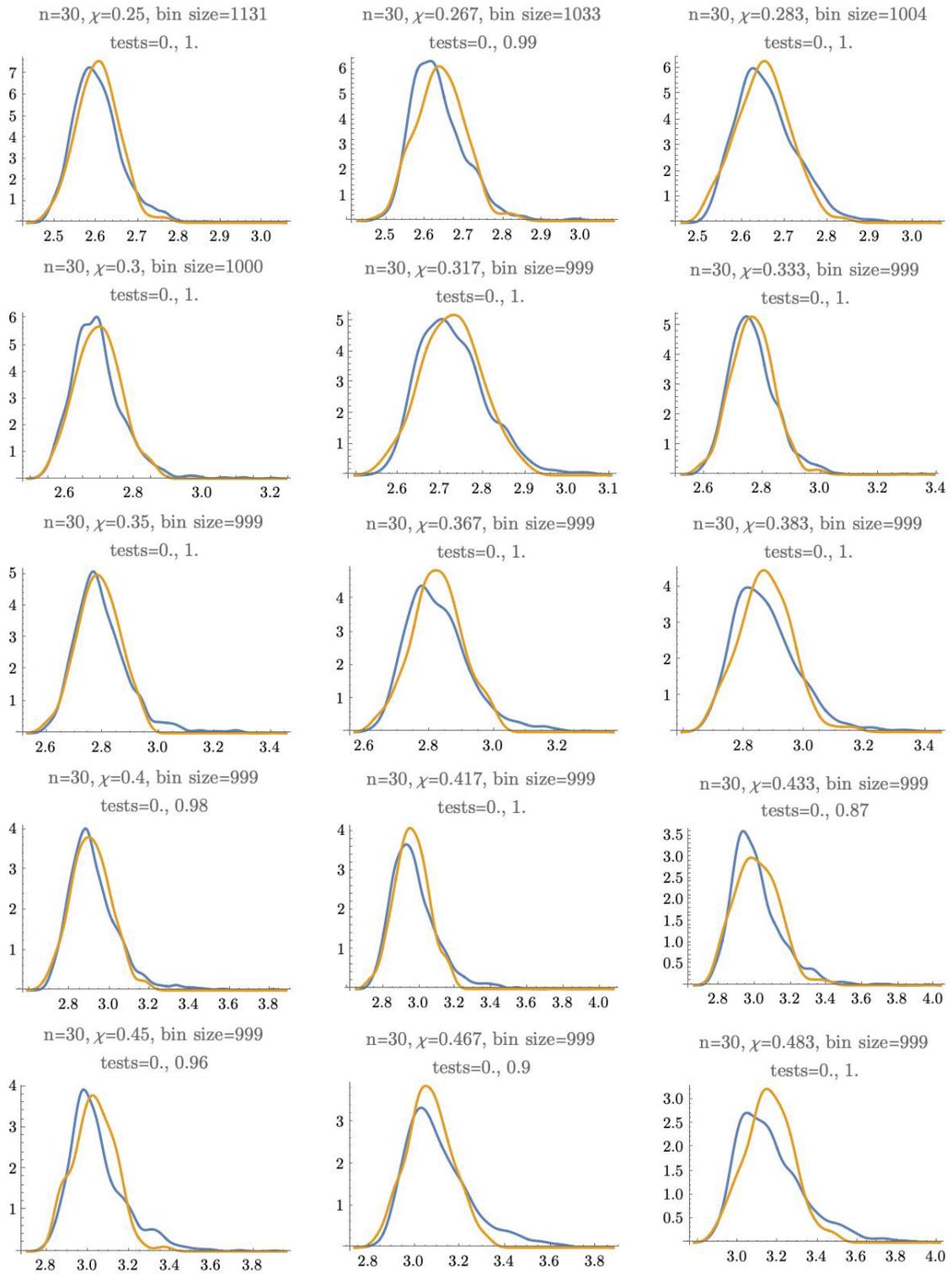

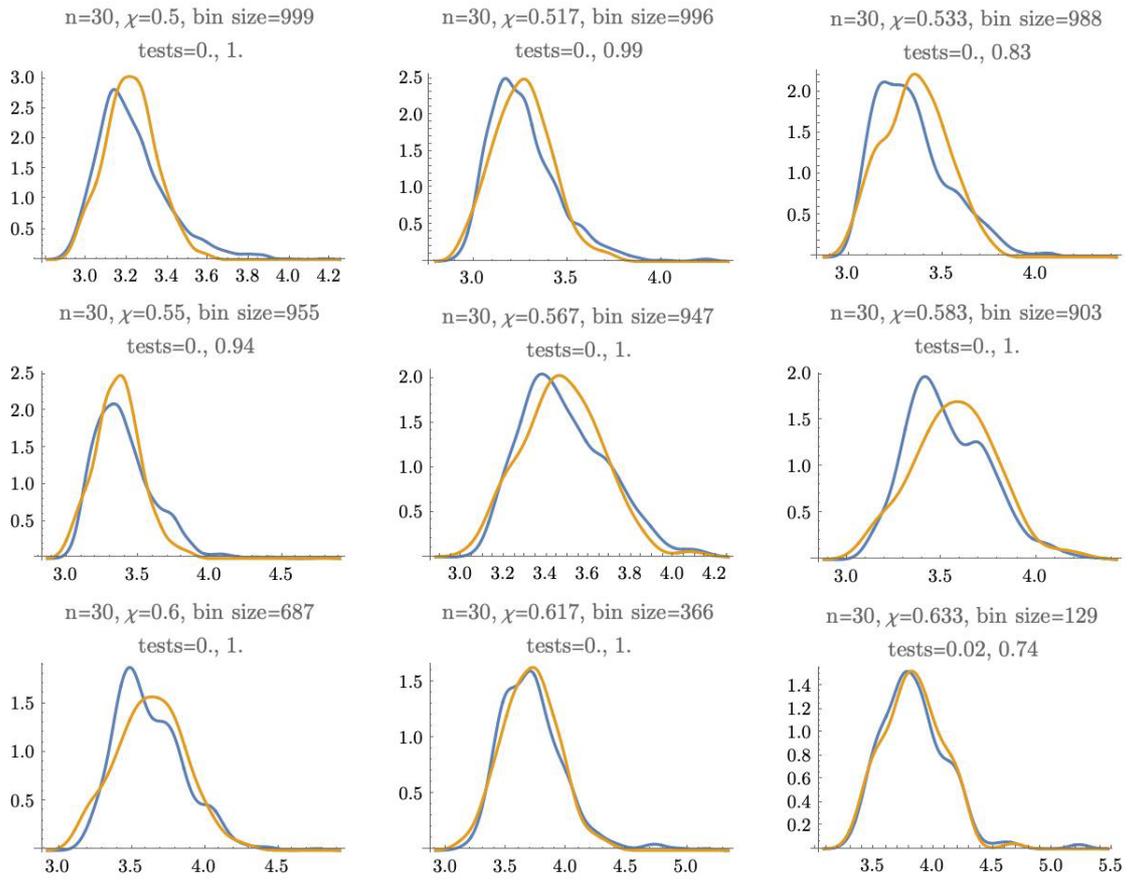

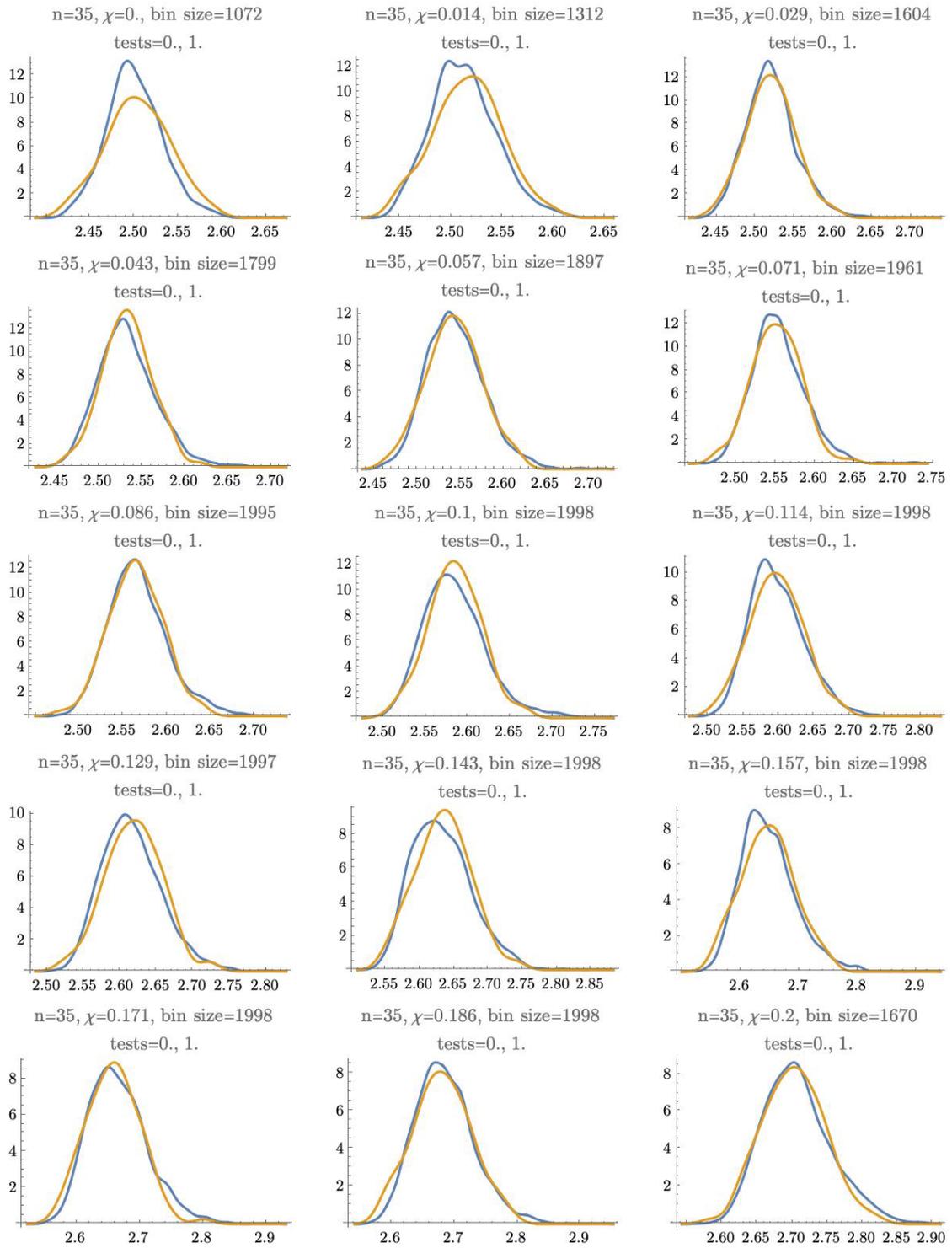

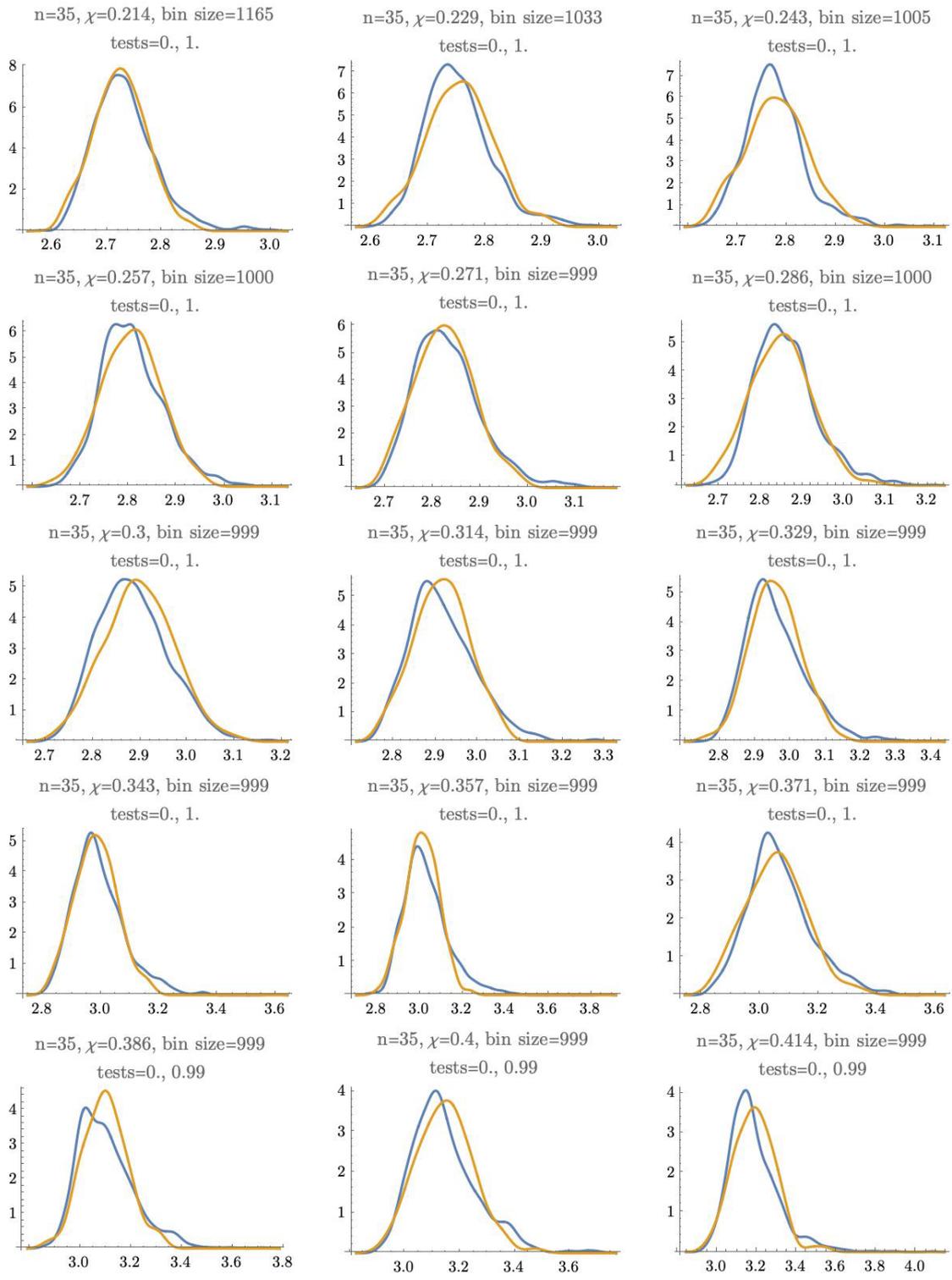

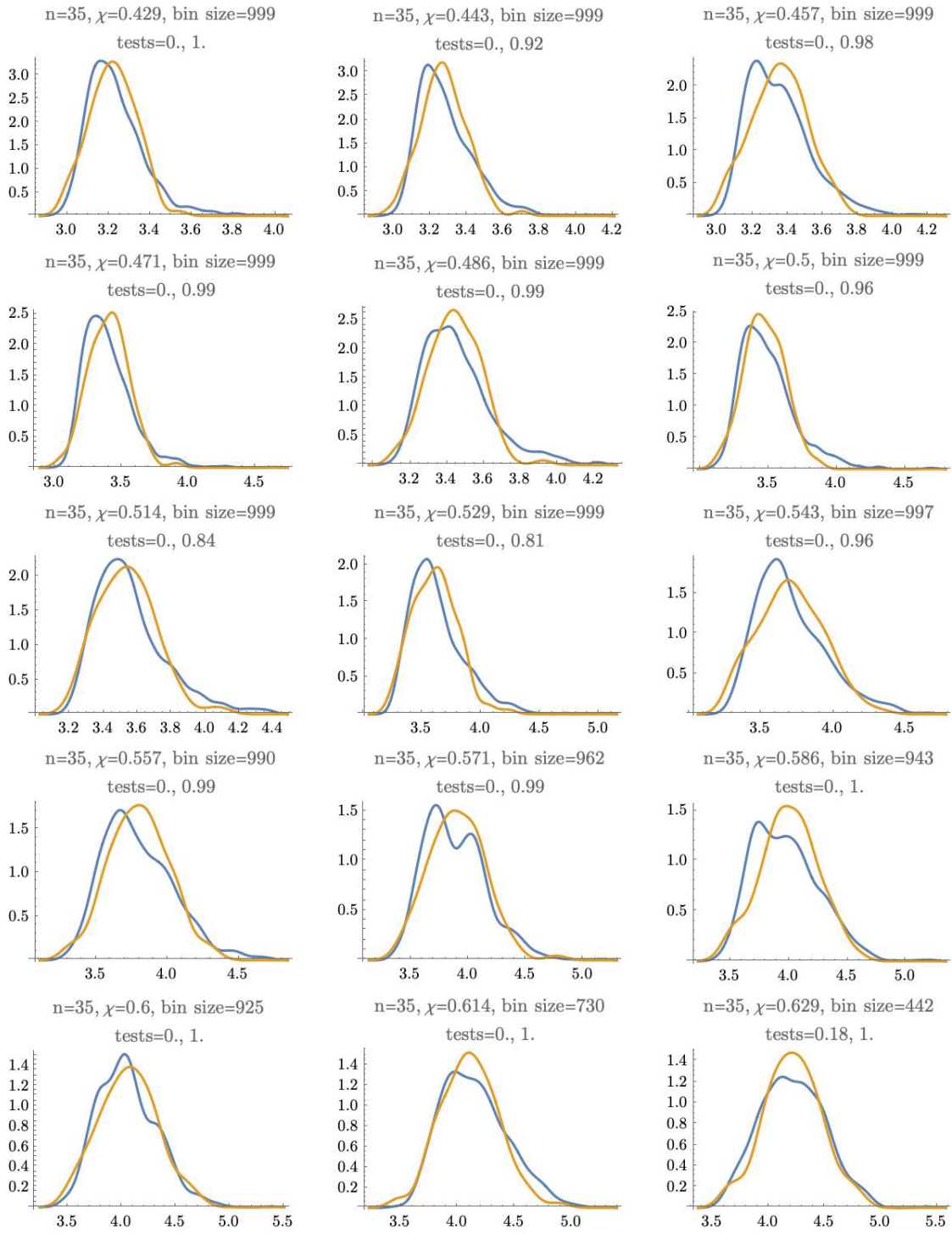

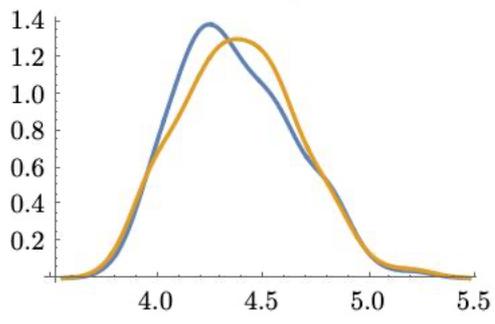

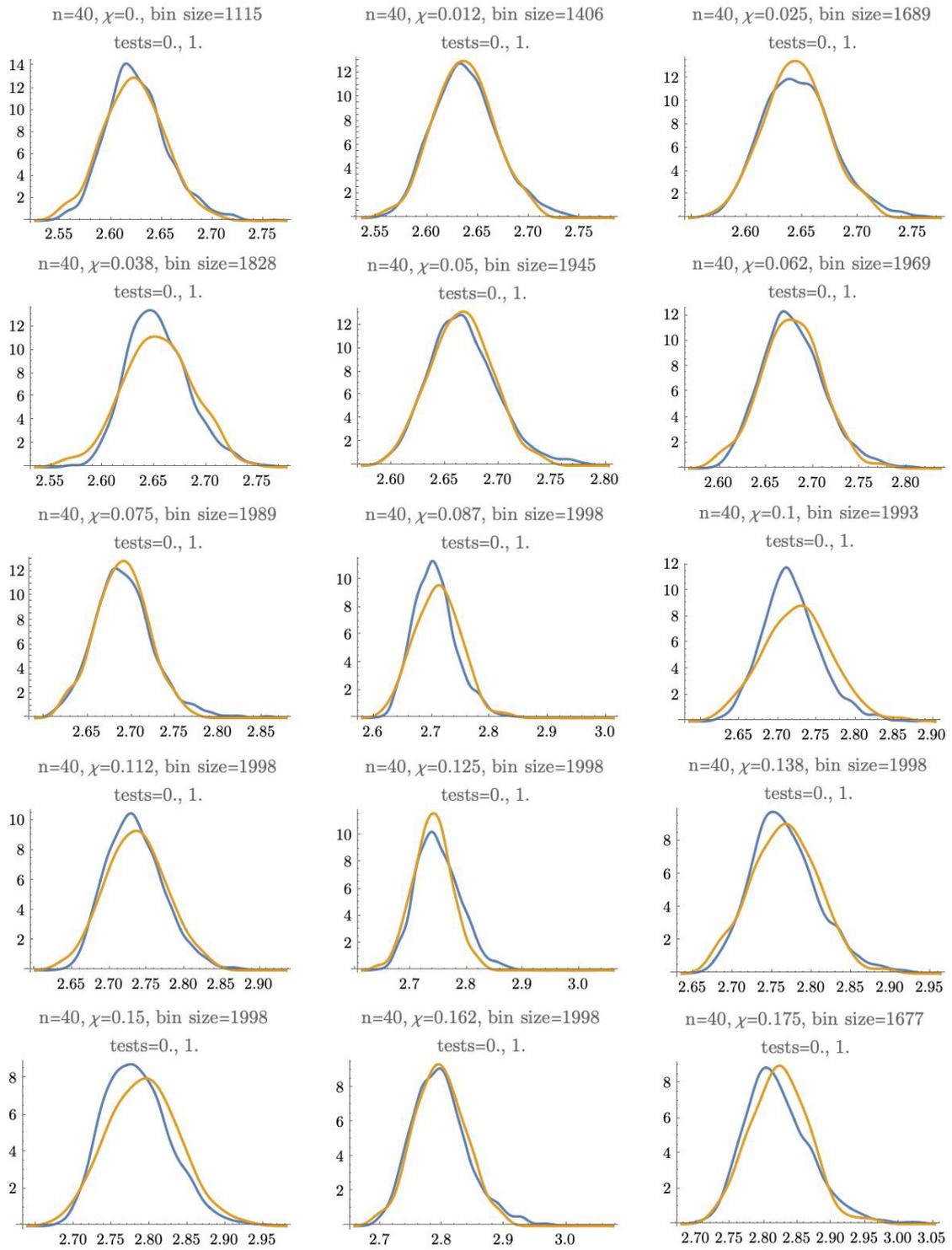

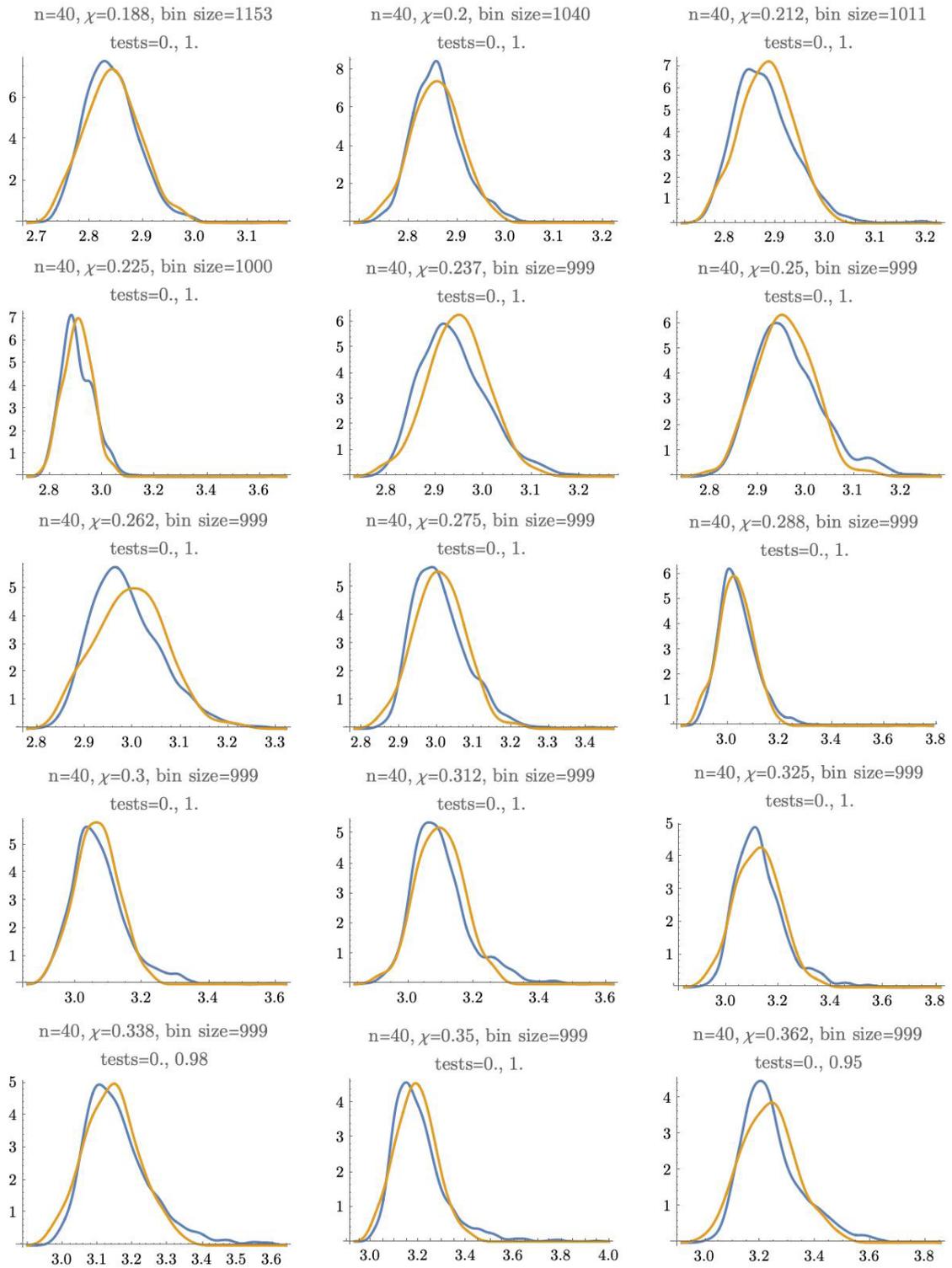

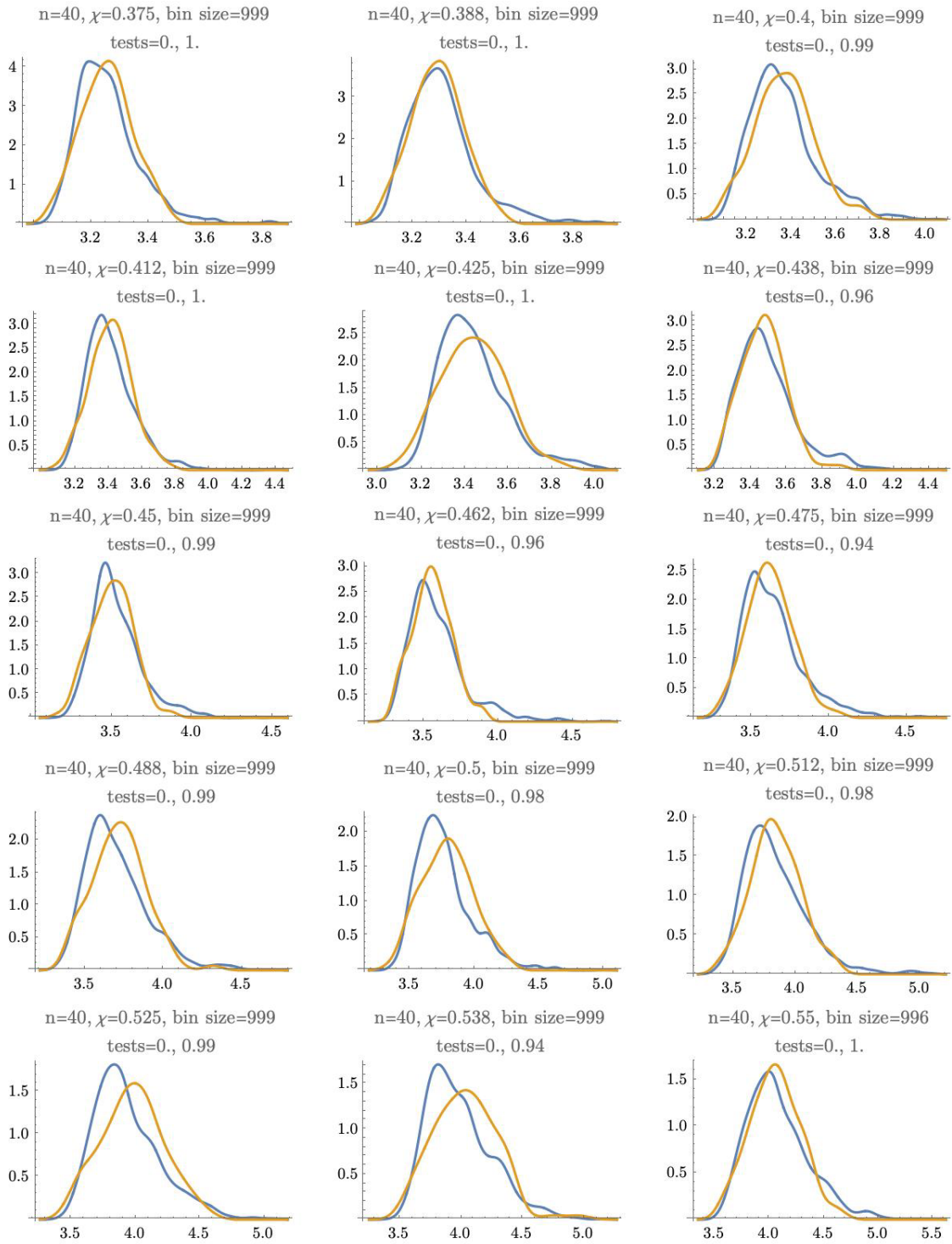

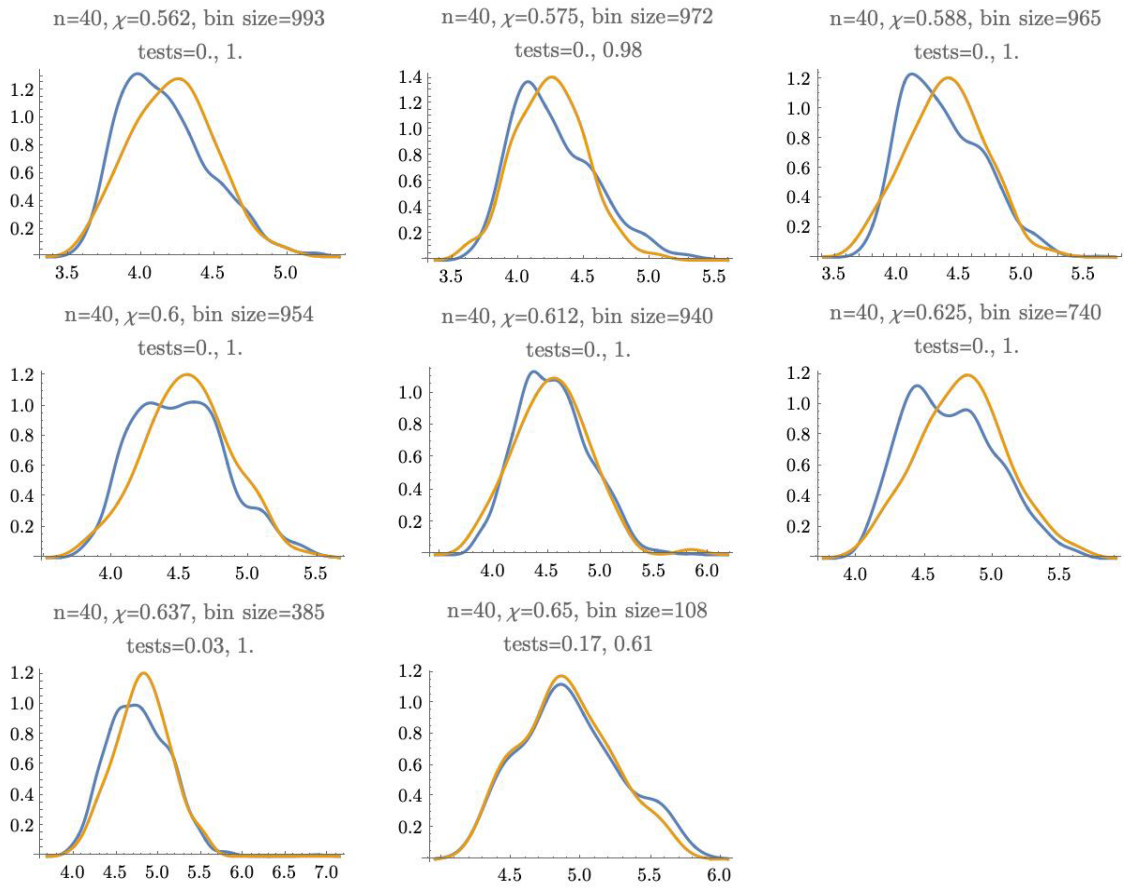

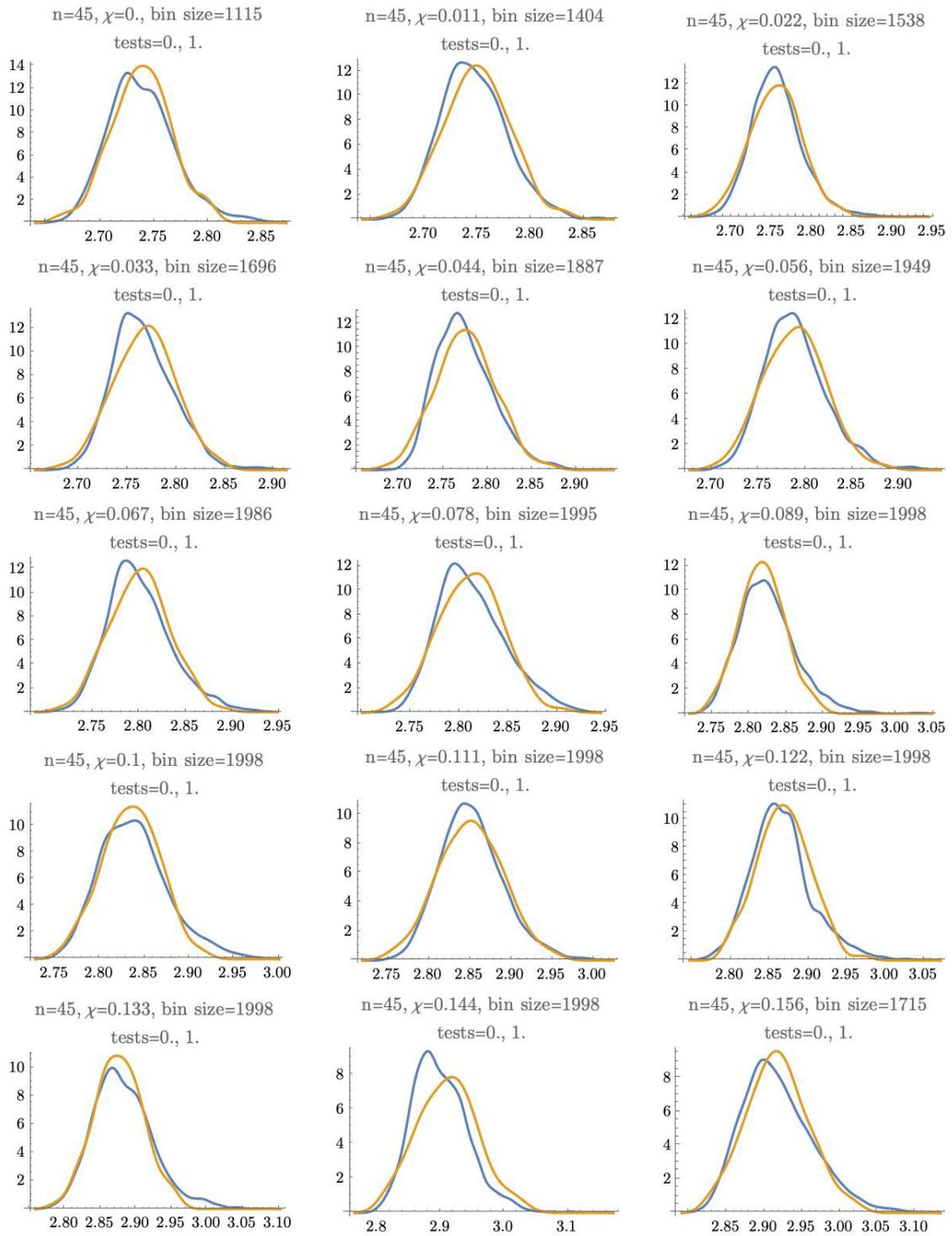

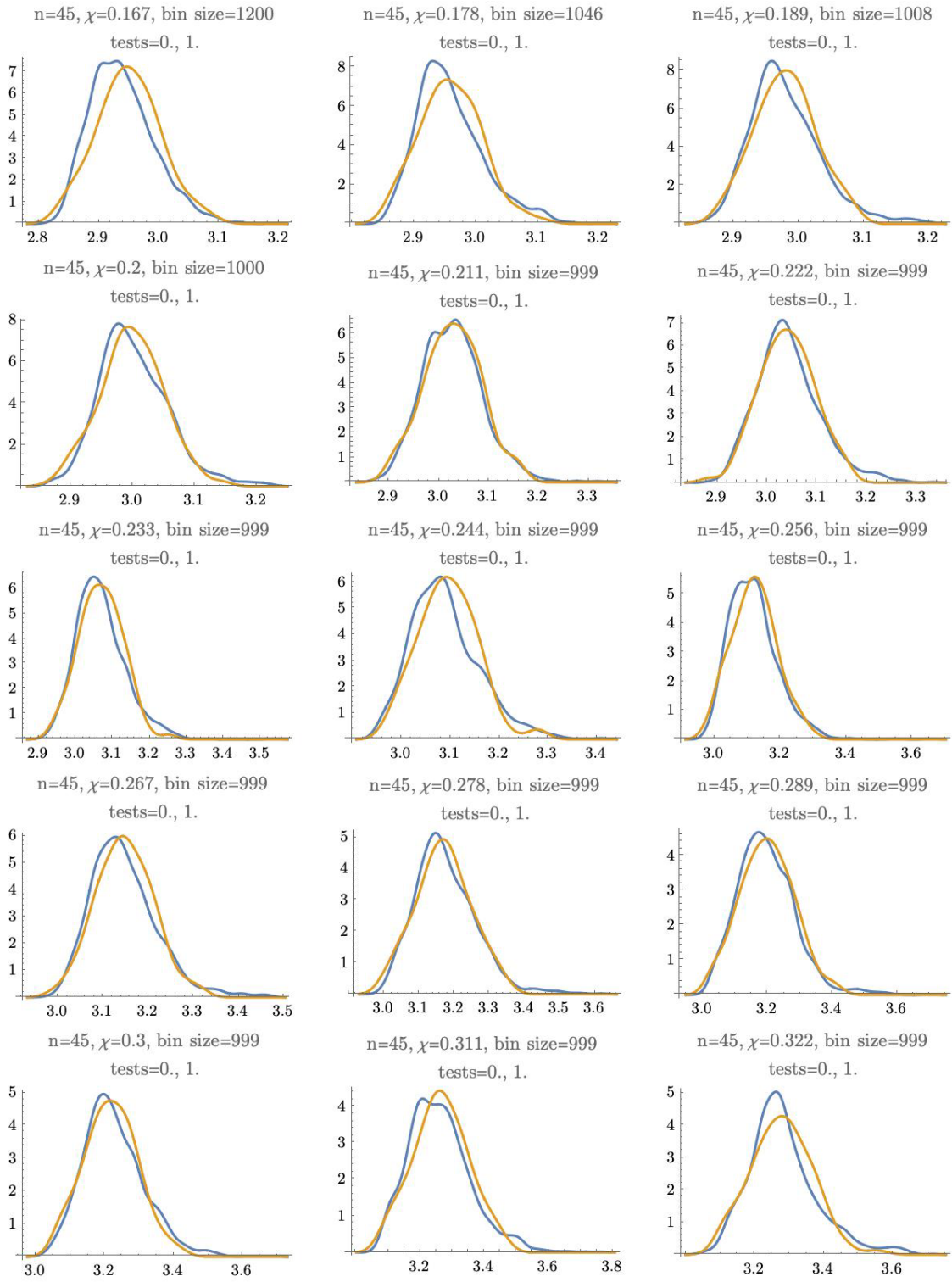

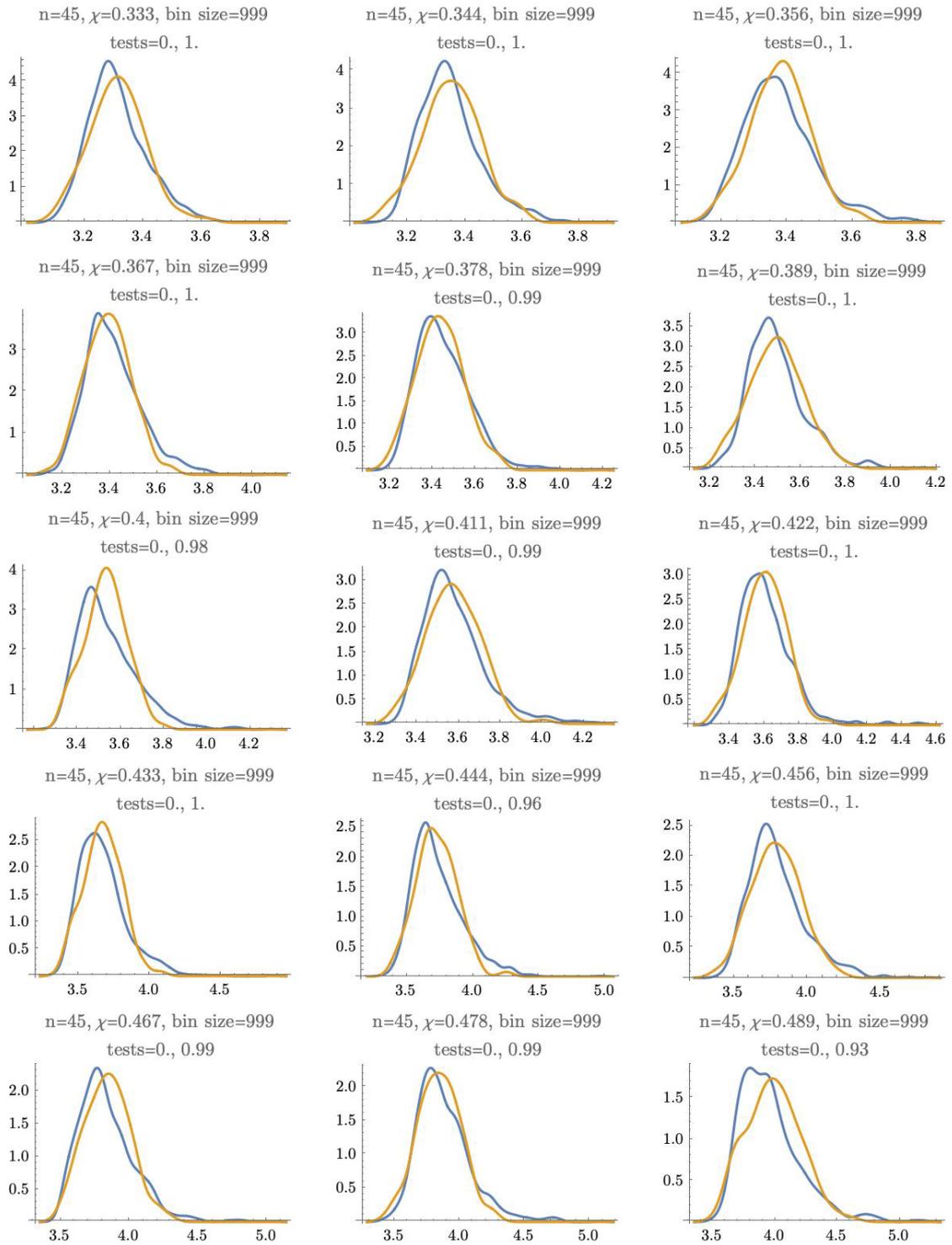

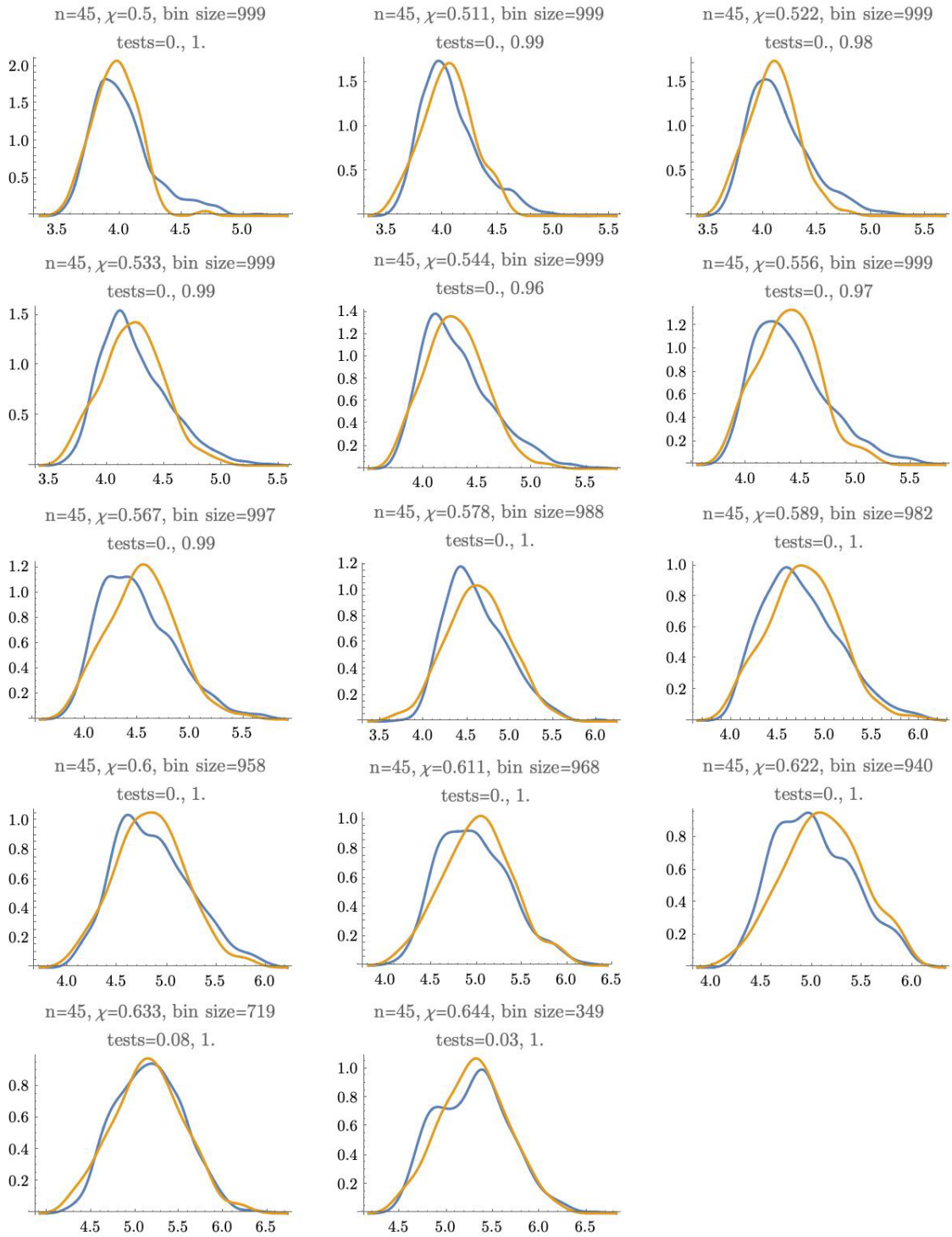

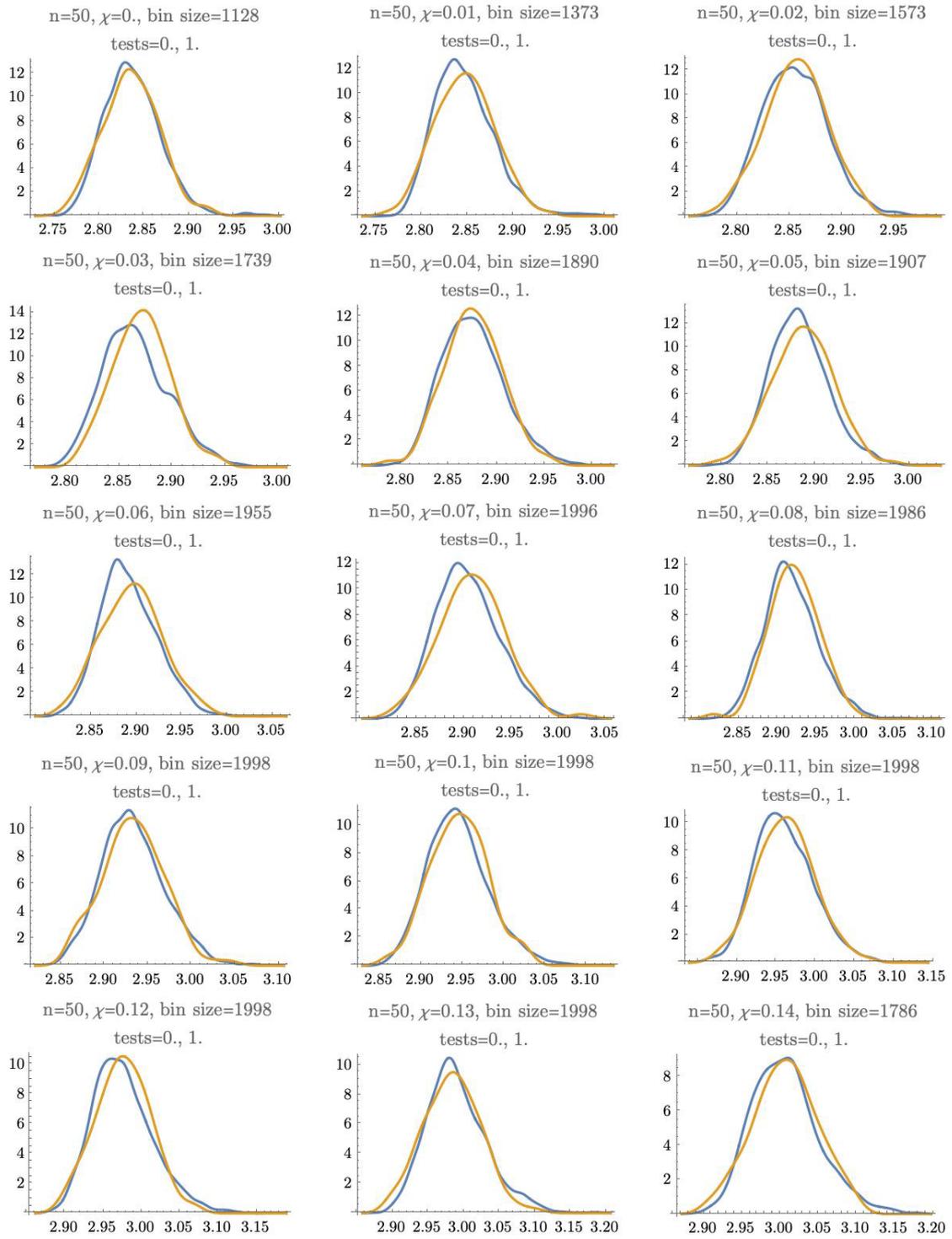

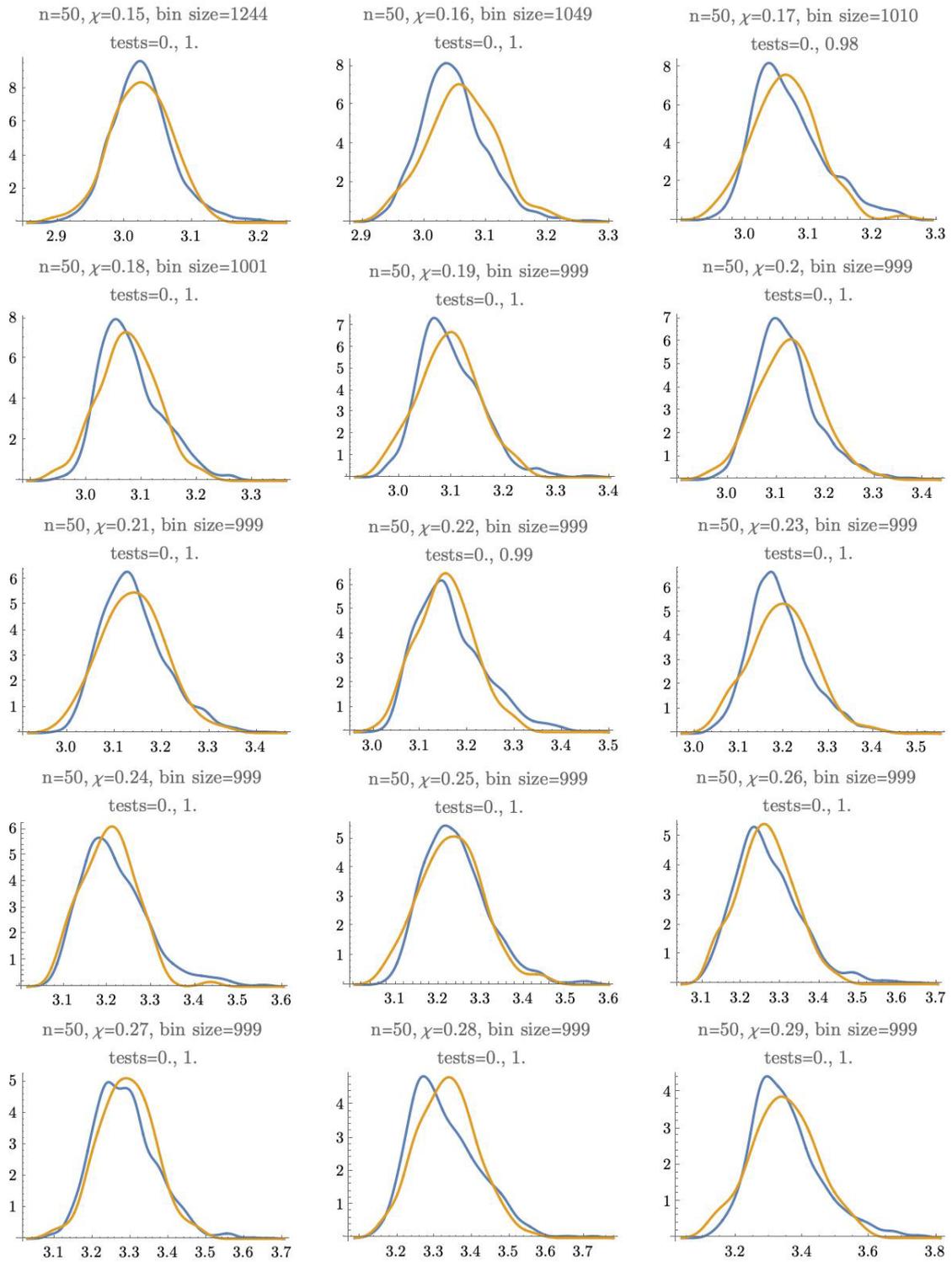

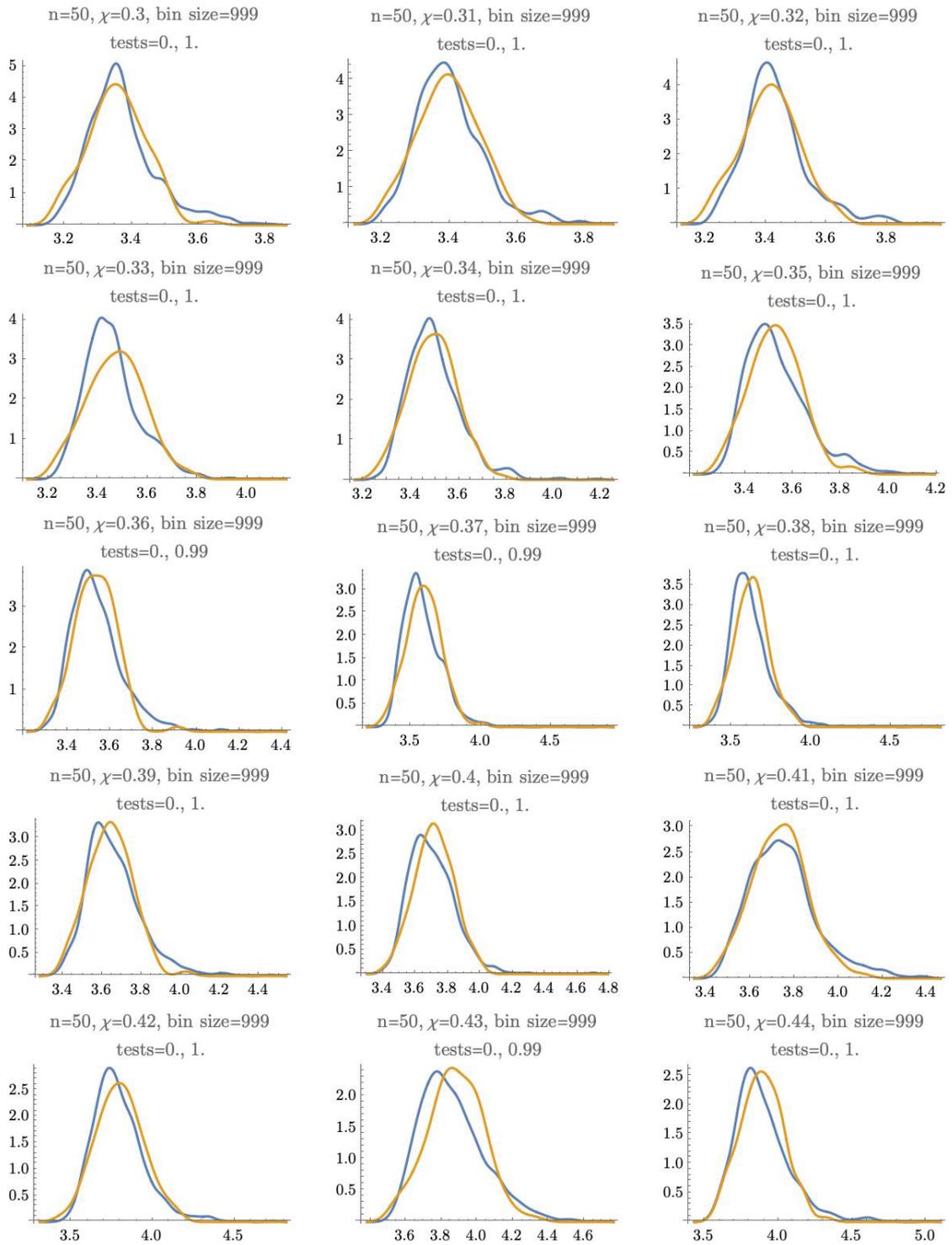

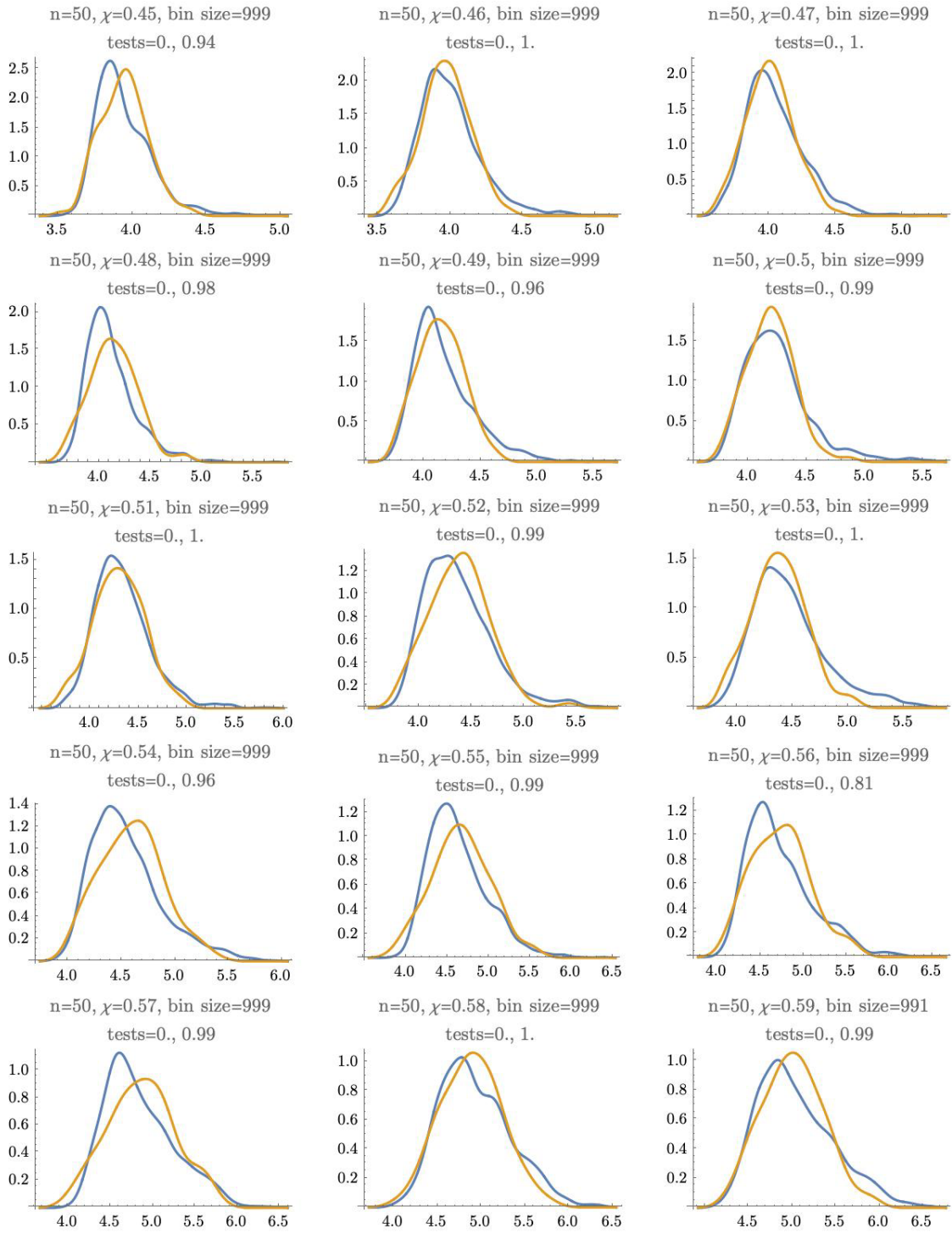

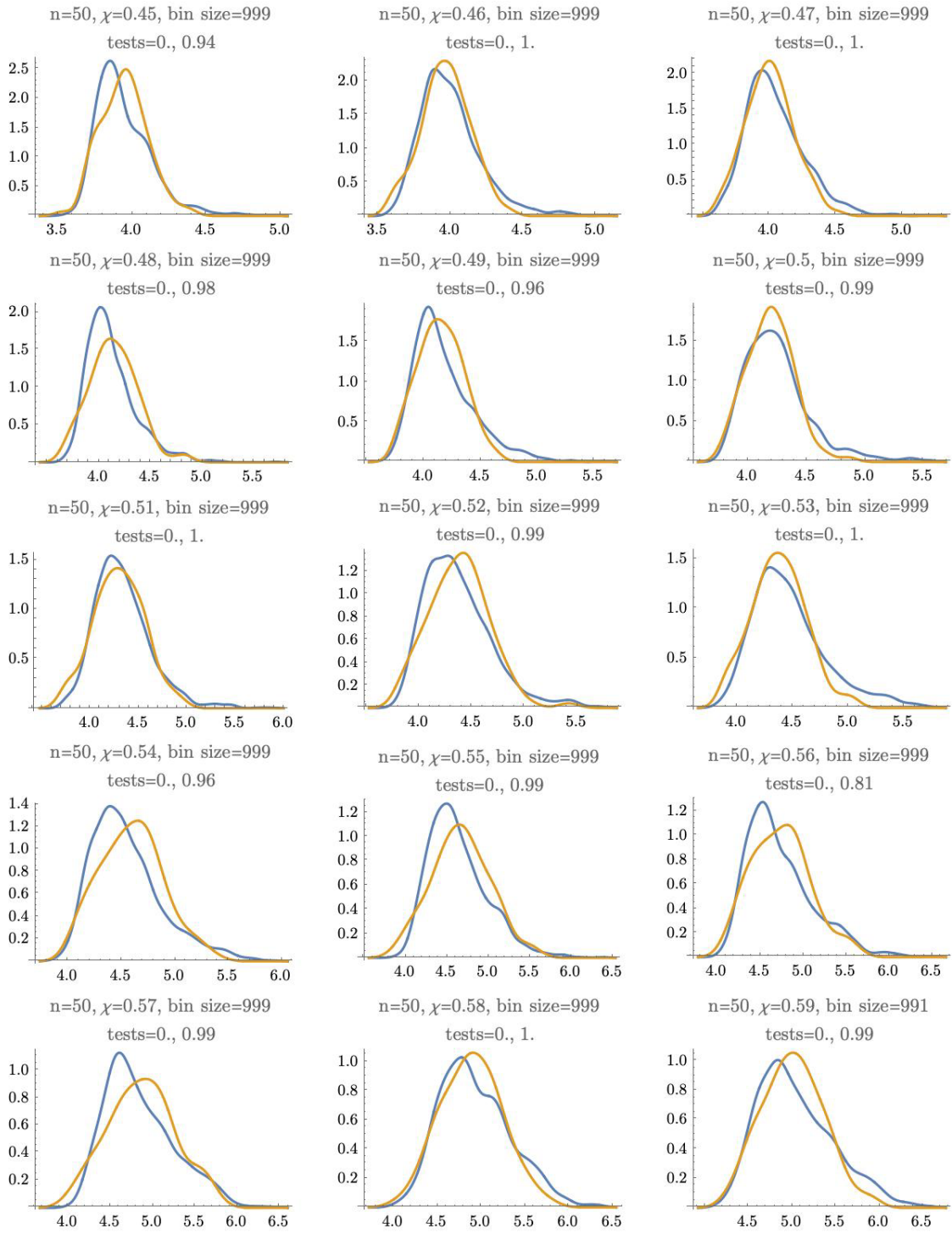

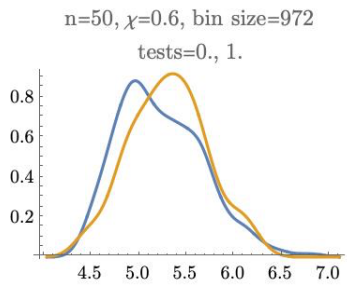

n=50, $\chi$=0.6, bin size=972
tests=0., 1.

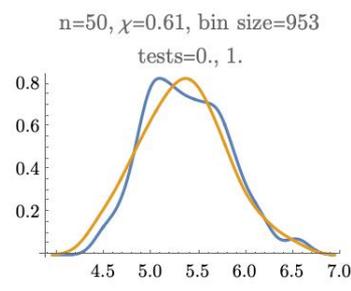

n=50, $\chi$=0.61, bin size=953
tests=0., 1.

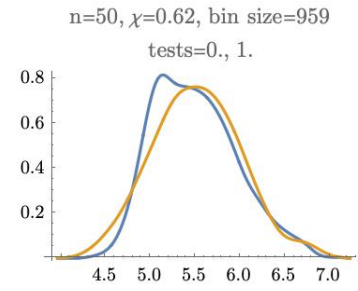

n=50, $\chi$=0.62, bin size=959
tests=0., 1.

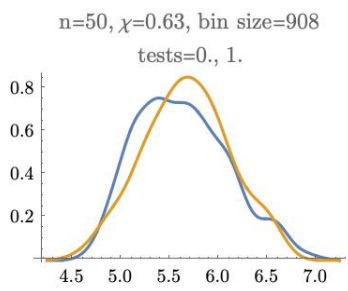

n=50, $\chi$=0.63, bin size=908
tests=0., 1.

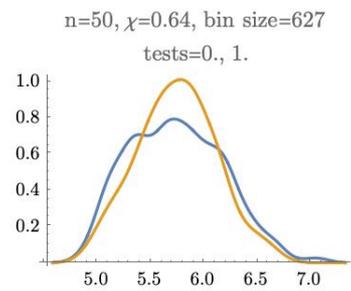

n=50, $\chi$=0.64, bin size=627
tests=0., 1.

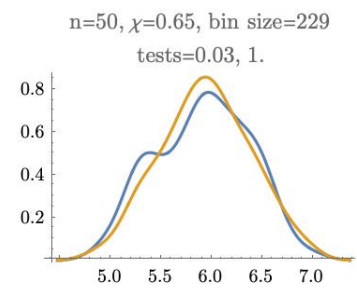

n=50, $\chi$=0.65, bin size=229
tests=0.03, 1.